\documentclass[aps,prl,twocolumn,superscriptaddress,numerical]{revtex4-2}
\usepackage[utf8]{inputenc}
\usepackage[english]{babel}
\usepackage[T1]{fontenc}
\usepackage{amsmath,amssymb,dsfont}
\usepackage[colorlinks,
citecolor=blue,
linkcolor=blue,
urlcolor=magenta]{hyperref}
\usepackage{tikz,cases,bm}

\usepackage{orcidlink}
\usepackage{multirow}
\usepackage{mathrsfs}
\usepackage[normalem]{ulem}

\DeclareMathAlphabet{\mathcal}{OMS}{cmsy}{m}{n}

\newcommand{\prlsection}[1]{\textbf{\textit{#1.}}}

\newcommand{\E}{\operatorname{erfc}}

\begin{document}
\title{Universal Tracer Statistics in Single-File Transport}

\author{Soumyabrata Saha~\orcidlink{0009-0006-7539-8801}}
\thanks{These authors contributed equally to this work.}
\affiliation{Department of Theoretical Physics, Tata Institute of Fundamental Research, Homi Bhabha Road, Mumbai 400005, India.}

\author{Jitendra Kethepalli~\orcidlink{0000-0001-7326-0231}}
\thanks{These authors contributed equally to this work.}
\affiliation{Laboratoire de Physique Th\'eorique et Mod\'elisation, CNRS UMR 8089, CY Cergy Paris Universit\'e, 95302 Cergy-Pontoise Cedex, France.}
\affiliation{JEIP, UAR 3573 CNRS, Collège de France, PSL Research University, 11 Place Marcelin Berthelot, 75321 Paris Cedex 05, France.}

\author{Benjamin Guiselin~\orcidlink{0000-0002-4663-2578}}
\affiliation{Laboratoire Charles Coulomb (L2C), Universit\'e de Montpellier, CNRS, Montpellier, France.}

\author{Jacopo De Nardis~\orcidlink{0000-0001-7326-0231}}
\affiliation{Laboratoire de Physique Th\'eorique et Mod\'elisation, CNRS UMR 8089, CY Cergy Paris Universit\'e, 95302 Cergy-Pontoise Cedex, France.}
\affiliation{JEIP, UAR 3573 CNRS, Collège de France, PSL Research University, 11 Place Marcelin Berthelot, 75321 Paris Cedex 05, France.}

\author{Tridib Sadhu~\orcidlink{0000-0003-0390-6257}}
\affiliation{Department of Theoretical Physics, Tata Institute of Fundamental Research, Homi Bhabha Road, Mumbai 400005, India.}

\begin{abstract}
We uncover an emergent universality in the large-scale, long-time statistics of a one-dimensional hard-rod gas evolving under two fundamentally different classes of microscopic dynamics: stochastic (diffusive) and unitary (ballistic). Remarkably, despite the difference of the two systems, the one-time joint distribution of the positions of multiple tracers exhibits identical non-Gaussian fluctuations, up to a simple dynamical scaling. This universality holds in both annealed and quenched ensembles, demonstrating a persistent memory of the initial state. Differences between the dynamics manifest at large scales only in multi-time statistics. Our conclusions are based on explicit large-deviation results for the one-time statistics of tracer pairs and the two-time statistics of a single tracer. Similar physics extends to current fluctuations, demonstrated explicitly in the quenched ensemble. We obtain these results from exact microscopic solutions for both dynamics and, independently, from fluctuating hydrodynamics in the ballistic case in the annealed ensemble. Our rare‑event simulations further corroborate these findings and provide a novel demonstration of sampling atypical fluctuations in both types of hard-rod gas.
\end{abstract}

\maketitle

Characterizing transport in interacting many-body systems is a central challenge in non-equilibrium physics. A prototypical setting arises in confined geometries, where particles move through narrow channels that forbid overtaking. This geometric constraint gives rise to \emph{single-file transport}, a distinctive form of collective dynamics observed across a wide range of physical systems \cite{2014_Flomenbom_Single,2017_Taloni_Single,2023_Cordes_Single}. Classical experimental realizations include biological channels \cite{1955_Hodgkin_The,1979_Urban_Ion}, colloidal assemblies \cite{2000_Wei_Single,2004_Lutz_Single,2005_Lin_From,2016_Locatelli_Single}, and porous media \cite{1996_Kulka_Nmr,1996_Hahn_Single,2009_Wang_Anomalous,2010_Das_Single,2010_Cambre_Experimental,2013_Hofling_Anomalous}, settings where dynamics occurs in the low-Reynolds-number regime and is effectively stochastic. More recently, cold-atom experiments have enabled the realization of quasi-one-dimensional hard-core quantum gases \cite{2004_Kinoshita_Observation,2004_Paredes_Tonks,2006_Kinoshita_Quantum,2024_Wienand_Emergence}, reviving interest in single-file systems with unitary dynamics.

Most theoretical efforts to understand the effects of the single-file constraint have focused on stochastic systems, such as lattice-based exclusion processes \cite{1983_Arratia_The,2015_Krapivksy_Tagged,2017_Imamura_Large,2018_Illien_Nonequilibrium,2022_Dandekar_Macroscopic,2022_Grabsch_Exact,2023_Grabsch_Driven,2025_Poncet_Full}, hard-core Brownian gases \cite{2010_Barkai_Diffusion,2014_Krapivsky_Large,2014_Hegde_Universal,2015_Sadhu_Large,2015_Krapivksy_Tagged,2017_Cividini_Tagged,2023_Dandekar_Dynamical,2025_Grabsch_Exact}, and even active matter \cite{2013_Illien_Active,2020_Dolai_Universal,2023_Rizkallah_Absolute,2023_Touzo_Interacting,2025_Akintunde_Single}. These systems exhibit a range of anomalous properties, including sub-diffusion \cite{1965_Harris_Diffusion,1974_Percus_Anomalous,1978_Alexander_Diffusion,1983_Arratia_The,2003_Kollmann_Single,2009_Barkai_Theory}, persistent memory \cite{2009_Derrida_Current,2013_Leibovich_Everlasting,2014_Lizana_Single,2015_Krapivsky_Dynamical,2015_Sadhu_Large,2021_Poncet_Cumulant}, and non-Gaussian fluctuations \cite{2009_Derrida_Current2,2013_Illien_Active,2014_Krapivsky_Large,2014_Meerson_Extreme,2023_Grabsch_Driven}. In parallel, unitary single-file systems, such as charged cellular automata \cite{2017_Medenjak_Diffusion,2022_Krajnik_Exact,2024_Krajnik_Universal,2025_Yoshimura_Anomalous} and quantum spin chains \cite{2019_Gopalakrishnan_Kinetic,2021_Pozsgay_Integrable,2026_Urilyon_Anomalous}, have also been found to display anomalous behaviour. This naturally raises the question of how sensitive these emergent properties are to the nature of microscopic dynamics.

A key probe of emergent properties in single-file systems is the statistics of tagged particles, commonly referred to as \emph{tracers}. While tracer statistics in stochastic systems have been extensively studied, most existing results capture only typical fluctuations \cite{1991_Majumdar_Two,2001_Rajesh_Exact,2015_Krapivsky_Dynamical,2016_Cividini_Correlation,2023_Dandekar_Dynamical}, with a full characterization available only in a few specific examples \cite{2014_Krapivsky_Large,2014_Hegde_Universal,2015_Sadhu_Large,2017_Imamura_Large,2021_Poncet_Cumulant,2023_Rana_Large}. For unitary dynamics, even less is known~\cite{2013_Roy_Tagged,2015_Roy_Tagged,2015_Sabhapandit_Exact}. A systematic comparative understanding of how microscopic dynamics govern tracer statistics, particularly large deviations, has so far remained elusive.

\begin{figure}[t]
\centering
\includegraphics[width=1.0\linewidth]{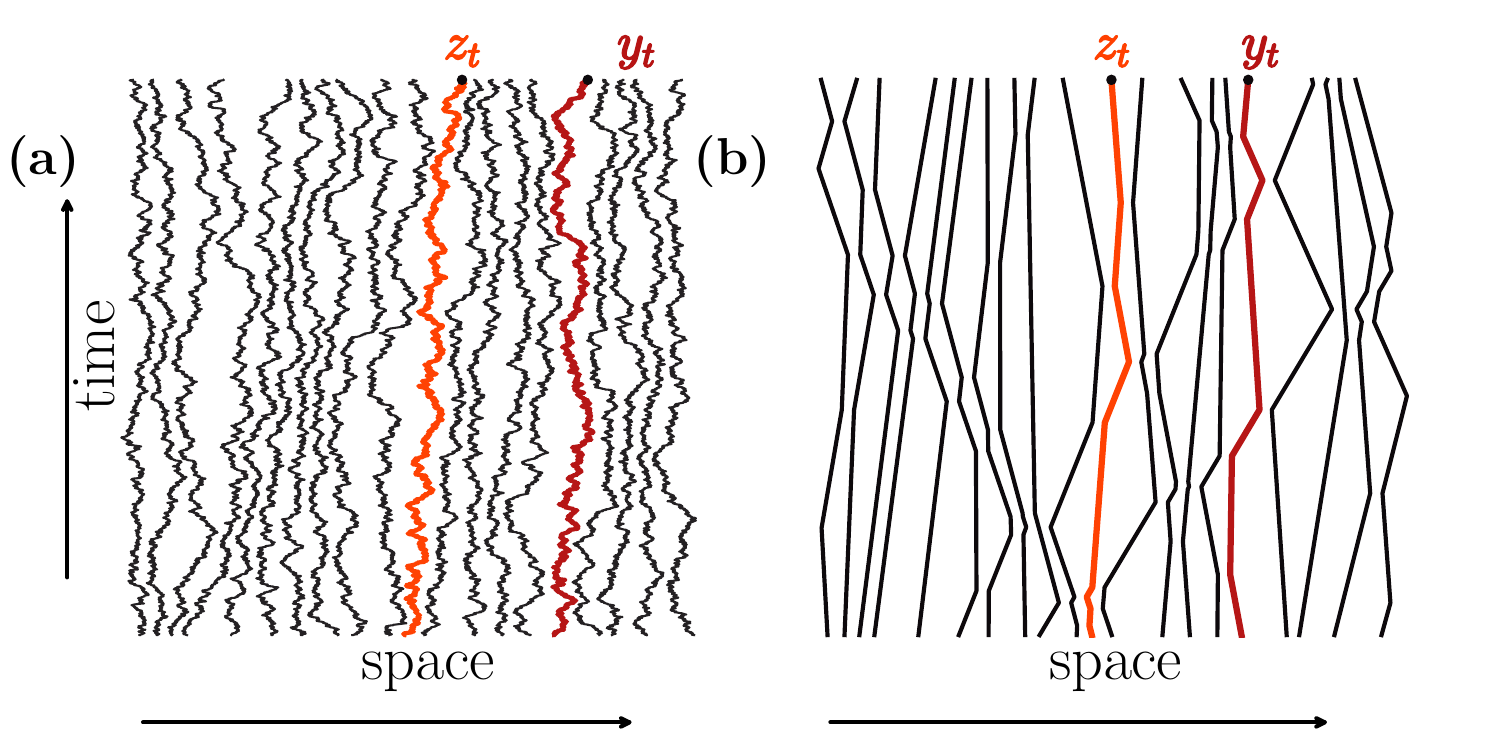}
\caption{Schematic trajectories of the (a) diffusive and (b) ballistic hard-rod gas. All rods have equal length $a$ and unit mass. In (a), particles undergo Brownian motion between collisions, while in (b) they move along straight lines governed by their velocities, which are exchanged upon collision. The trajectories of two tracers are highlighted, with their positions at time $t$ denoted by $z_t$ and $y_t$.}
\label{fig:schematic_trajectories}
\end{figure}

To address this gap, we study the \emph{hard-rod gas} \cite{1969_Percus_Exact,1983_Boldrighini_One,1991_Spohn_Large,2025_Kundu_Ballistic}, which provides a minimal interacting model realizing the two distinct classes of single-file dynamics (see Fig.~\ref{fig:schematic_trajectories}). In the stochastic setting, particles diffuse according to Brownian motion subject to excluded-volume constraints, whereas in the unitary version they move ballistically and exchange velocities upon elastic collision. We refer to these as the \emph{diffusive} and \emph{ballistic} hard-rod gases, respectively. Despite their simplicity, both versions capture a broad range of anomalous single-file behaviour \cite{1965_Jepsen_Dynamics,1973_Levitt_Dynamics,1967_Lebowitz_Kinetic,2003_Kollmann_Single,2008_Lizana_Single,2024_Powdel_Conserved,2025_Chahal_Stochastic,2025_Grabsch_Exact} and serves as a natural continuum counterpart of lattice exclusion models \cite{1968_Macdonald_Kinetics,1969_Macdonald_Concerning,2004_Schonherr_Exclusion,2005_Schonherr_Hard,2007_Derrida_Non,2013_Krapivsky_Dynamics,2015_Mallick_The,2025_Derrida_Les_Houches}.

In this \emph{Letter}, we present a comparative characterization of the tracer statistics in the two hard-rod gases, focusing on the large-deviation statistics of the one-time position of a tracer and of tracer pairs as well as the two-time position of a single tracer. Our explicit results reveal an emergent universality in one-time statistics across the two classes of dynamics, up to a simple dynamical scaling. Interestingly, the statistics of the gap between tracers reaches stationarity in which even this scaling difference disappears. This universality holds irrespective of whether the initial conditions are \emph{annealed} or \emph{quenched}, despite the well-known \cite{2009_Derrida_Current2,2012_Krapivksy_Fluctuations,2013_Leibovich_Everlasting,2015_Krapivksy_Tagged,2015_Sadhu_Large,2014_Lizana_Single} persistent-memory effects in single-file systems. Similar conclusions extend to current fluctuations, for which we obtain explicit large-deviation results in the quenched initial state, a particularly challenging problem in interacting-particle systems \cite{2009_Derrida_Current,2015_Sadhu_Large,2017_Imamura_Large,2021_Poncet_Cumulant,2022_Mallick_Exact,2023_Rana_Large,2025_Sharma_large}. 

The emergent universality is especially striking in view of the markedly different large-scale descriptions of the two gases. The diffusive gas is governed by the conventional fluctuating hydrodynamics \cite{2007_Derrida_Non,2025_Grabsch_Macroscopic,2026_Saha_Bottom} with a single locally conserved field, whereas the ballistic gas possesses an extensive set of local conserved quantities requiring a generalized hydrodynamics (GHD) description \cite{1991_Spohn_Large,2016_Bertini_Transport,2016_Castro_Emergent}. Yet differences between the two dynamics manifest at large scales only in multi‑time statistics. Our results are established via two independent methods: exact microscopic solutions for both gases, and a hydrodynamic approach for the ballistic gas. They are further corroborated by importance-sampling simulations \cite{1989_Ferrenberg_Optimized,2014_Hartmann_High,2025_Hartmann_Numerical} applied here (see End Matter) for the first time to interacting hard-rod gases in both stochastic and unitary limits.

\prlsection{Setup} A crucial difference between the two gases lies in the source of fluctuations. For the diffusive gas, fluctuations in the tracer position arise from the stochastic dynamics and the randomness in the initial particle positions. In contrast, for the ballistic gas they originate solely from the randomness in the initial velocities and positions of particles. For a comparative analysis, we treat the randomness in the initial velocities of the ballistic gas on the same footing as the dynamical noise in the diffusive gas, and distinguish both from the randomness in the initial positions. This distinction is useful in single-file systems, which are known \cite{2009_Derrida_Current2,2012_Krapivksy_Fluctuations,2013_Leibovich_Everlasting,2015_Krapivksy_Tagged,2015_Sadhu_Large,2014_Lizana_Single} to retain a memory of initial conditions. By analogy with disordered systems, this motivates defining, for a generic observable $\mathcal{O}$, Annealed ($\mathcal{A}$) and Quenched ($\mathcal{Q}$) cumulant-generating functions (CGFs)
\begin{equation}\label{eq:cgf}
\mu_{\mathcal{A}}(\lambda)=\ln\langle\mathrm{e}^{\lambda\mathcal{O}}\rangle_{\text{d}+\text{p}},\quad\mu_{\mathcal{Q}}(\lambda)=\langle\ln\langle\mathrm{e}^{\lambda\mathcal{O}}\rangle_{\text{d}}\rangle_{\text{p}},
\end{equation}
where the subscripts `d' and `p' denote averaging over dynamical fluctuations and initial positions, respectively.

The initial positions in both gases are sampled from a uniform density profile $\bar{\rho}$, with the tracer placed at the origin. In the ballistic case, the rod velocities are independently drawn from a Gaussian distribution with variance $2D_0$, a canonical choice \cite{2008_Rigol_Thermalization,2024_Singh_Thermalization} for the generalized-Gibbs ensemble (GGE). Here, $D_0$ is chosen to match the bare diffusion coefficient of the diffusive gas, enabling a direct quantitative comparison between the two gases.

\begin{figure*}
\centering
\includegraphics[width=1.0\linewidth]{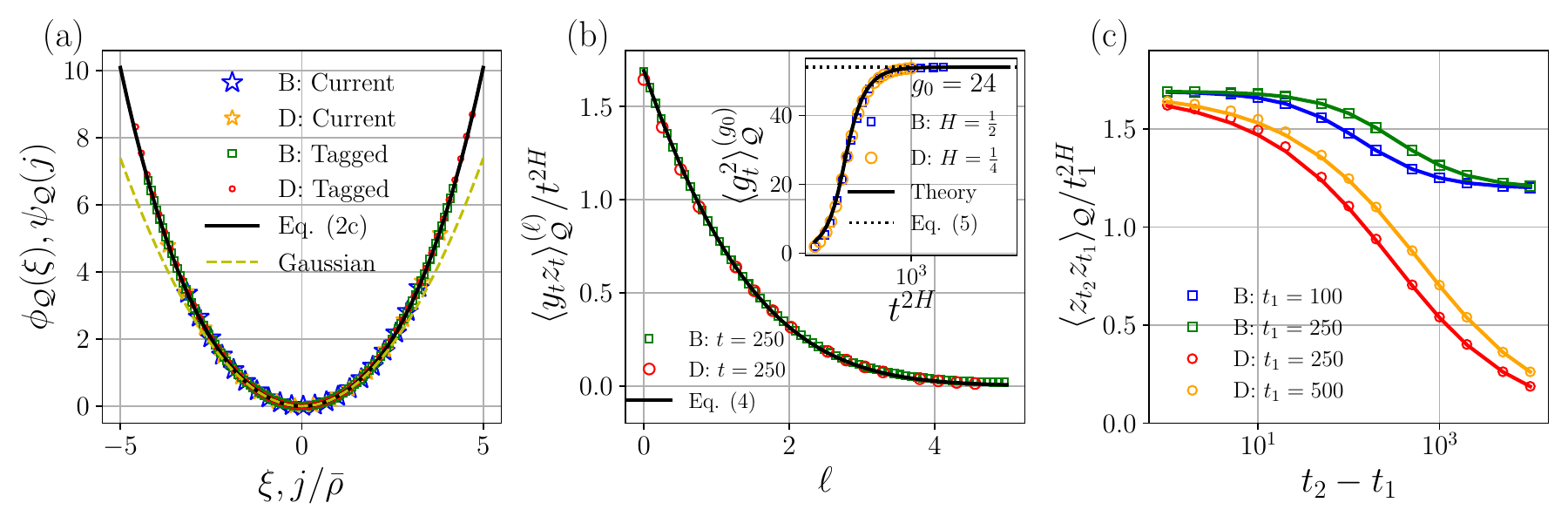}
\caption{Results for both gases in the quenched ensemble with $D_0=0.5$, $a=1$, and $\bar{\rho}=0.25$. Markers (D: diffusive, B: ballistic) are numerical results; solid lines are analytical predictions. Plot (a) shows the tracer LDF \eqref{eq:1_time_1_tracer_quench_ldf} and the current LDF, confirming the relation $\psi_{\mathcal{Q}}\left(j\right)=\phi_{\mathcal{Q}}\left(\tfrac{j}{\bar{\rho}}\right)$. The dashed line indicates a Gaussian LDF with the same variance, highlighting the non-Gaussian tails. Plot (b) shows the equal-time covariance of tracer pairs \eqref{eq:corr_2tracer_1time} as a function of the initial rescaled gap $\ell$, with the inset illustrating the approach of the gap variance to its stationary value \eqref{eq:stationary_separation}. In both (a) and (b), the two gases yield identical statistics, confirming the universality. Plot (c) shows the two-time covariances \eqref{eq:2time_cov_quench} of a single tracer, where the universality breaks down, with signatures of aging.
}
\label{fig:quench_numerics}
\end{figure*}

\prlsection{Results} We show that the long-time distribution of the tracer position $z_t$ follows a large-deviation form
\begin{subequations}\label{eq:large_dev_one_time_one_tracer}
\begin{equation}\label{eq:large_dev_1tracer_1time}
\Pr\bigg(\frac{z_t}{t^{2H}}=\xi\bigg)\asymp\mathrm{e}^{-{t^{2H}}\phi(\xi)},
\end{equation}
where the exponent $H$ governs the scale of fluctuations: $H=\tfrac{1}{4}$ for the diffusive case and $H=\tfrac{1}{2}$ for the ballistic case. Strikingly, the large-deviation function (LDF) $\phi(\xi)$ is identical for the two gases. It is obtained as the Legendre transform \cite{2007_Derrida_Non,2009_Touchette_The} of the corresponding scaled-CGF $t^{-2H}\mu(\lambda)$ in \eqref{eq:cgf}, and takes distinct forms in the two ensembles.
In the annealed ensemble,
\begin{equation}\label{eq:1_time_1_tracer_anneal_ldf}
\phi_{\mathcal{A}}(\xi)=\bar{r}\Big(\sqrt{I(\xi)}-\sqrt{I(-\xi)}\Big)^2,
\end{equation}
whereas in the quenched ensemble the LDF is expressed in parametric form as
\begin{equation}\label{eq:1_time_1_tracer_quench_ldf}
\phi_{\mathcal{Q}}(\xi)=-\bar{r}\sum_{\varepsilon=\pm} \int_{\varepsilon\xi}^{\infty}\mathrm{d}u\ln\bigg\{1+\frac{\mathcal{B}^{\varepsilon}}{2}\E\Big(\frac{u}{\sqrt{4 D_0}}\Big)\bigg\},
\end{equation}
\end{subequations}
with $I(\xi)=\tfrac{1}{2}\int_{\xi}^{\infty}\mathrm{d}u\,\E\left(\tfrac{u}{\sqrt{4D_0}}\right)$, $\E$ the complementary error function, $\mathcal{B}^{\pm}=\mathrm{e}^{\pm B}-1$, and the parameter $B$ determined by the condition $\partial_B\phi_{\mathcal{Q}}(\xi)=0$.

Note how the rod length $a$ enters the LDFs solely through the renormalized density $\bar{r}=\tfrac{\bar{\rho}}{1-a\bar{\rho}}$. This non-trivial feature does not follow immediately \cite{2023_Rizkallah_Duality} from simple excluded-volume arguments \cite{1936_Tonks_The,1968_Lebowitz_Time}; for instance, the annealed current LDF \cite{2024_Grabsch_From,2026_Kethepalli_Ballistic} does not have this property.

Several consequences follow immediately from the identical LDFs of the two gases. For both systems, the variances in the two ensembles are related \cite{2012_Krapivksy_Fluctuations,2015_Krapivksy_Tagged} by $\left<z_t^2\right>_{\mathcal{A}}=\sqrt{2}\left<z_t^2\right>_{\mathcal{Q}}$. Non-Gaussian tails are evident from the large-$\xi$ asymptotics: $\phi_{\mathcal{A}}(\xi)\sim|\xi|$ and $\phi_{\mathcal{Q}}(\xi)\sim|\xi|^3$, consistent with earlier observations \cite{2009_Derrida_Current2,2014_Meerson_Extreme,2014_Krapivsky_Large} for diffusive systems. In the annealed ensemble, our results also recover known \cite{2024_Grabsch_From} fourth cumulant of the tracer.

A particularly non-trivial consequence concerns the distribution of time-integrated current $n_t$ across the origin, which has a large-deviation form analogous to \eqref{eq:large_dev_1tracer_1time}. The corresponding quenched LDF follows directly from the relation $\psi_{\mathcal{Q}}\left(j\right)=\phi_{\mathcal{Q}}\left(\tfrac{j}{\bar{\rho}}\right)$ via the single-file constraint \cite{2015_Sadhu_Large,Supp_Mat}. Numerical evidence supporting this relation for both gases is shown in Fig.~\ref{fig:quench_numerics}a. In stark contrast, analogous results for exclusion processes remain out of reach \cite{2009_Derrida_Current2,2021_Poncet_Cumulant,2023_Rana_Large}.

The universality further extends to the \emph{two-tracer statistics}. Placing a second tracer at an initial gap $g_0$ from the first tracer at the origin, the joint distribution of their positions at long times takes a large-deviation form
\begin{subequations}\label{eq:large_dev_one_time_two_tracer}
\begin{equation}\label{eq:large_dev_2tracers_1time}
\Pr\bigg(\frac{z_t}{t^{2H}}=\xi,\frac{y_t}{t^{2H}}=\zeta\bigg)\asymp\mathrm{e}^{-t^{2H}\Phi^{(\ell)}(\xi,\zeta)}
\end{equation}
where the superscript denotes the rescaled initial gap, $\ell=\tfrac{g_0}{t^{2H}}$. The LDF in the annealed ensemble is
\begin{align}\label{eq:2tracer_annealed}
&\Phi^{(\ell)}_{\mathcal{A}}(\xi,\zeta)=-\bar{r}\sum_{\varepsilon=\pm}\Big\{\mathcal{B}_1^{\varepsilon}I(\varepsilon\xi)+\mathcal{B}_2^{\varepsilon}I(\varepsilon(\zeta-\ell))\Big\}\nonumber\\
&+\mathcal{B}_1^{+}\mathcal{B}_2^{+}I\big(\zeta-a\bar{r}\ell_{\text{r}}\big)+\mathcal{B}_1^{-}\mathcal{B}_2^{-}I\big(-\xi+\ell_{\text{r}}\big),
\end{align}
\end{subequations}
with $\ell_{\text{r}}=\ell(1-a\bar{\rho})$. The corresponding expression for the quenched ensemble is provided in \cite{Supp_Mat}.

An immediate corollary of \eqref{eq:2tracer_annealed} is the two-tracer covariance
\begin{equation}\label{eq:corr_2tracer_1time}
\big<y_tz_t\big>_{\mathcal{A}}^{(\ell_{\mathcal{A}})}=\sqrt{2}\big<y_tz_t\big>_{\mathcal{Q}}^{(\ell_{\mathcal{Q}})}\simeq t^{2H}\frac{2}{\bar{r}}I\big(\ell_{\text{r}}\big),
\end{equation}
where the superscript represents the rescaled initial gaps $\ell_{\mathcal{A}}=\ell$ and $\ell_{\mathcal{Q}}=\sqrt{2}\ell$ for the corresponding ensembles. The first equality is a generalization of the annealed--quenched correspondence for the tracer variance \cite{2012_Krapivksy_Fluctuations,2015_Krapivksy_Tagged}, while the final expression is verified numerically in Fig.~\ref{fig:quench_numerics}b. Two notable conclusions follow from \eqref{eq:corr_2tracer_1time}. First, the rod length $a$ effectively renormalize the rescaled initial gap $\ell\to\ell_{\text{r}}$. Second, the variance of the gap $g_t=y_t-z_t$ reaches a stationary value (see Fig.~\ref{fig:quench_numerics}b)
\begin{equation}\label{eq:stationary_separation}
\big<g_t^2\big>_{\mathcal{A}}^{(g_{\mathcal{A}})}=\sqrt{2}\big<g_t^2\big>_{\mathcal{Q}}^{(g_{\mathcal{Q}})}\simeq\frac{2}{\bar{r}(1+a\bar{r})}g_0,
\end{equation}
where the superscripts $g_{\mathcal{A}}=g_0$ and $g_{\mathcal{Q}}=\sqrt{2} g_0$ denote the initial gap between the tracers for the corresponding ensemble. Notably, the universality fully manifests in the stationary gap statistics, where even the time-rescaling drops out.

The difference between the two gases emerges only in the \emph{multi-time statistics}. For example, the joint distribution of the position of a single tracer at two large times $t_1$ and $t_2\equiv t$ takes a large-deviation form
\begin{subequations}\label{eq:large_dev_two_time_one_tracer}
\begin{equation}\label{eq:large_dev_1tracer_2times}
\Pr\bigg(\frac{z_{t_1}}{t^{2H}}=\xi_1,\frac{z_{t_2}}{t^{2H}}=\xi_2\bigg)\asymp\mathrm{e}^{-t^{2H}\varphi^{(\tau_1,\tau_2)}(\xi_1,\xi_2)}
\end{equation}
where $\tau_j=t_j/t$, and with the LDF $\varphi$ distinct for the two gases. In the annealed ensemble, the LDFs can be written together as
\begin{align}\label{eq:1tracer_annealed_2time}
&\varphi_\mathcal{A}^{(\tau_1,\tau_2)}(\xi_1,\xi_2)=-\bar{r}\sum_{\varepsilon=\pm}\Bigg\{\sum_{j=1}^2\mathcal{B}_j^{\varepsilon}\tau_j^{2H}I\bigg(\varepsilon\frac{\xi_j}{\tau_j^{2H}}\bigg)\nonumber\\
&+\prod_{j=1}^2\mathcal{B}_j^{\varepsilon}\int_{\varepsilon\xi_j}^{\infty}\mathrm{d}u_j\,\mathcal{H}_{\tau_1,\tau_2}(u_1,u_2)\mathcal{G}_{\tau_2-\tau_1}(u_2|u_1)\Bigg\},
\end{align}
\end{subequations}
with the difference in the two dynamics appearing in the function $\mathcal{H}_{\tau_1,\tau_2}(u_1,u_2)=\tfrac{1}{2}\E\left(\tfrac{u_1}{2\sqrt{D_0\tau_1}}\right)$ for the diffusive case and $\mathcal{H}_{\tau_1,\tau_2}(u_1, u_2)=\Theta\left(\tfrac{u_2}{\tau_2}-\tfrac{u_1}{\tau_1}\right)$ for the ballistic gas, and $\mathcal{G}_{\tau}(u_2|u_1)=\left(4\pi D_0\tau^{4H}\right)^{-1/2}\mathrm{e}^{-\tfrac{(u_2-u_1)^2}{4D_0\tau^{4H}}}$. The corresponding quenched results are presented in \cite{Supp_Mat}.

The difference between the two gases is best illustrated in the two-time covariance of tracer position, which for the annealed ensemble, follows directly from \eqref{eq:1tracer_annealed_2time},
\begin{equation}\label{eq:2time_cov_anneal}
\big<z_{t_1}z_{t_2}\big>_{\mathcal{A}}\simeq\frac{\sqrt{D_0}}{\sqrt{\pi}\bar{r}}\big(t_1^{2H}+t_2^{2H}-|t_1-t_2|^{2H}\big).
\end{equation}
This shows that the typical process for the tracer \cite{2015_Krapivsky_Dynamical,2023_Dandekar_Dynamical} is a fractional-Brownian motion \cite{1968_Mandelbrot_Fractional}, with the two gases belonging to different classes distinguished by $H$. The corresponding quenched result \cite{Supp_Mat},
\begin{equation}\label{eq:2time_cov_quench}
\big<z_{t_1}z_{t_2}\big>_{\mathcal{Q}}\simeq\frac{\sqrt{D_0}}{\sqrt{\pi}\bar{r}}\bigg(\sqrt{t_1^{4H}+t_2^{4H}}-|t_1-t_2|^{2H}\bigg)
\end{equation}
exhibits aging \cite{2008_Henkel_Local,2011_Berthier_Theoretical}. Similar results hold for the current, with $\left<n_{t_1}n_{t_2}\right>=\bar{\rho}^2\left<z_{t_1}z_{t_2}\right>$ in both ensembles \cite{Supp_Mat}. 

The contrast between \eqref{eq:2time_cov_anneal} and \eqref{eq:2time_cov_quench} reflects the persistence of memory even for the stationary initial state. The quenched result \eqref{eq:2time_cov_quench} is confirmed numerically in Fig.~\ref{fig:quench_numerics}c. The annealed result \eqref{eq:2time_cov_anneal} can be indirectly confirmed \cite{Supp_Mat,1997_Krug_Persistence,2015_Krapivsky_Dynamical,2023_Dandekar_Dynamical} from \eqref{eq:2time_cov_quench} by time-translation $t\to T+t$ and taking $T\to\infty$.

In the rest of the \emph{Letter}, we explain how the above results are obtained independently using microscopic and hydrodynamic approaches.

\prlsection{Microscopics} Let $\mathbf{x}_0=\{x_0^{(i)}\}$ and $\mathbf{x}_t=\{x_t^{(i)}\}$ denote the initial and final positions (at time $t$) of the hard-rods, where $i$ labels the particle rank. For both hard-rod gases, the joint propagator admits a Bethe-ansatz form \cite{Supp_Mat} (see also \cite{1998_Rodenbeck_Calculating,2009_Lizana_Diffusion,2015_Krapivksy_Tagged,2017_Cividini_Tagged})
\begin{equation}\label{eq:joint_prop_point_part}
P(\mathbf{x}_t|\mathbf{x}_0)=\sum_{\sigma}\prod_i\mathcal{G}_t\big(x_t^{(i)}-ia\,\big|\,x_0^{(\sigma_i)}-\sigma_ia\big),
\end{equation}
where $\mathcal{G}_t(x|x_0)$ is the single-particle Gaussian propagator with variance $2D_0t^{4H}$ defined earlier, and $\sigma$ denotes permutations of the index set.

The tracer distribution is obtained \cite{Supp_Mat} by marginalizing over the final positions of the remaining particles, followed by averaging over initial positions according to the ensembles discussed in \eqref{eq:cgf}. The resulting distribution yields the large-deviation asymptotics announced in (\ref{eq:large_dev_one_time_one_tracer},~\ref{eq:large_dev_one_time_two_tracer},~\ref{eq:large_dev_two_time_one_tracer}). These microscopic calculations are rendered \cite{Supp_Mat} more tractable by mapping the hard-rod gas to an equivalent point-particle representation via the coordinate transformation \cite{1936_Tonks_The,1968_Lebowitz_Time} $X_t^{(i)}=x_t^{(i)}-ia$.

\prlsection{Hydrodynamics} Our microscopic analyses rely on integrable structures that are often not generalizable. This motivates hydrodynamic approaches formulated in terms of the relevant slow modes, leading to the development of macroscopic fluctuation theory (MFT) \cite{2001_Bertini_Fluctuations,2002_Bertini_Macroscopic,2005_Bertini_Current,2006_Bertini_Non,2007_Derrida_Non,2007_Tailleur_Mapping,2008_Tailleur_Mapping,2009_Derrida_Current2,2015_Bertini_Macroscopic,2022_Mallick_Exact,2024_Saha_Large,2025_Derrida_Les_Houches} for diffusive systems, and more recently its ballistic counterpart, ballistic macroscopic fluctuation theory (BMFT) \cite{2023_Doyon_Emergence,2023_Doyon_Ballistic,2025_Yoshimura_Anomalous,2025_Yoshimura_Hydrodynamic,2026_Kethepalli_Ballistic}. While MFT has been extensively used to characterize tracer statistics \cite{2014_Krapivsky_Large,2015_Krapivksy_Tagged,2023_Dandekar_Dynamical,2024_Grabsch_Tracer,2025_Berlioz_Tracer,2025_Grabsch_Exact}, the corresponding application of BMFT remains scarce \cite{2025_Chahal_Stochastic}. In what follows, we present a non-trivial application by recovering the tracer statistics (\ref{eq:large_dev_one_time_one_tracer},~\ref{eq:large_dev_one_time_two_tracer},~\ref{eq:large_dev_two_time_one_tracer}) for the annealed ensemble in the ballistic case, thereby providing an essential check of BMFT in the characterization of rare fluctuations.

In BMFT, all large-scale fluctuations originate from the initial state and are propagated by deterministic Euler-scale hydrodynamics. For ballistic hard-rods, the Euler-scale evolution is described by GHD \cite{2016_Bertini_Transport,2016_Castro_Emergent} (see also \cite{2017_Doyon_Large,2018_Doyon_Soliton,2018_Nardis_Hydrodynamic,2020_Ruggiero_Quantum,2019_Bastianello_Generalized,2025_Hubner_Diffusive,2025_Doyon_Generalized}), which governs the evolution of the single-particle phase-space density $\varrho_t(x,\theta)$ on the position($x$)-velocity($\theta$) plane. The probability of the density $\varrho_t(x,\theta)$ at time $t> 0$ is then given by a path integral
\begin{equation}\label{eq:bmft_trajectory_prob}
\Pr[\varrho_t]=\int\mathcal{D}[\varrho_s]\delta\Big(\partial_s\varrho_s+\partial_{x}\big(v^{\rm eff}_s\varrho_s\big)\Big)\mathcal{P}[\varrho_0].
\end{equation}
Here the delta functional enforces the GHD equation for hard-rods with effective velocity $v^{\rm eff}_t(x,\theta)=\tfrac{\theta-a\int\mathrm{d}\theta\,\theta\varrho_t(x,\theta)}{1-a\int\mathrm{d}\theta\,\varrho_t(x,\theta)}$, representing the quasiparticle velocity and incorporating the effects of scattering shifts~\cite{1983_Boldrighini_One,1991_Spohn_Large,2026_Kethepalli_Ballistic}. The fluctuations of $\varrho_0$ in the initial state obey a large-deviation form $\mathcal{P}[\varrho_0]\asymp\mathrm{e}^{-\mathcal{F}_{\rm B}[\varrho_0]}$, with generalized-Gibbs free energy \cite{2007_Derrida_Non,2026_Kethepalli_Ballistic}
\begin{equation}
\mathcal{F}_{\rm B}[\varrho_0]=\int\mathrm{d}x\,\mathrm{d}\theta\Big\{\mathfrak{n}_0\ln\Big(\frac{\mathfrak{n}_0}{\bar{\mathfrak{n}}_0}\Big)-(\mathfrak{n}_0-\bar{\mathfrak{n}}_0)\Big\},
\end{equation}
where $\mathfrak{n}_0(x,\theta)=\tfrac{\varrho_0(x,\theta)}{1-a\int\mathrm{d}\theta\,\varrho_0(x,\theta)}$ is the normal-mode density. This quantifies the cost of creating a profile that deviates from $\bar{\mathfrak{n}}_0$ associated to typical profile $\bar{\varrho}_0$.

The statistics of generic macroscopic observables expressible in terms of the density profile, such as the tracer position, can be computed using \eqref{eq:bmft_trajectory_prob}. The tracer position $z_t\equiv z_t[\varrho_t]$ is implicitly defined as a functional of the phase-space density via the single-file constraint~\cite{2015_Krapivksy_Tagged}
\begin{equation}\label{eq:zt def}
\int_{0}^{z_t}\mathrm{d}x\,\rho_t(x)=\int_{-\infty}^{0} dx \big(\rho_0(x)-\rho_t(x)\big),
\end{equation}
where the number density $\rho_t(x)=\int\mathrm{d}\theta\,\varrho_t(x,\theta)$. The generating function $\left<\mathrm{e}^{\lambda z_t}\right>=\int\mathcal{D}[\varrho,\hat{\varrho}]\,\mathrm{e}^{\lambda z_t[\varrho_t]-\mathcal{S}_{\rm B}[\varrho,\hat{\varrho}]}$ is then governed by the BMFT-action $\mathcal{S}_{\rm B}=\mathcal{F}_{\rm B}+\mathcal{L}_{\rm B}$ with
\begin{equation}\label{eq:action_B}
\mathcal{L}_{\rm B}[\varrho,\hat{\varrho}]=\int_0^t\mathrm{d}s\,\int \mathrm{d}x\,\mathrm{d}\theta\,\hat{\varrho}_s\big(\partial_s\varrho_s+\partial_{x}(v^{\rm eff}_s\varrho_s)\big)
\end{equation}
where the conjugate field $\hat{\rho}_s$ is introduced using Fourier representation \cite{2026_Kethepalli_Ballistic} of the delta functional in \eqref{eq:bmft_trajectory_prob}.

Under a ballistic space-time rescaling, the action is extensive in $t\left(\gg1\right)$, which enables a saddle-point evaluation \cite{Supp_Mat,2026_Kethepalli_Ballistic} of the large-time asymptotics of the generating function. The corresponding large-deviation asymptotics is then obtained by a Legendre transform \cite{Supp_Mat} of the scaled-CGF, reproducing the result in \eqref{eq:1_time_1_tracer_anneal_ldf}. Although solving the saddle-point equations is challenging, they become tractable using the mapping to point particles described earlier. This mapping for the hydrodynamic variables reads $\varrho_t(x,\theta)\to r_t(X,\theta)=n_0(x,\theta)$ with
\begin{equation}\label{eq:ghd_mapping}
x[X,r_t]=X+\frac{a}{2}\int\mathrm{d}X'\,r_t(X')\operatorname{sgn}(X-X'),
\end{equation}
which reduces the interacting problem to an analytically tractable one. Details of the calculations are discussed in \cite{Supp_Mat}. The same framework extends to yield \cite{Supp_Mat} the two-tracer \eqref{eq:2tracer_annealed} and two-time \eqref{eq:1tracer_annealed_2time} statistics.

The corresponding analysis for diffusive hard-rods within MFT are far more challenging due to the additional fluctuations from the inherent stochastic dynamics. They are formally incorporated within the MFT framework, which describes the evolution of the number density $\rho_t(x)$. The analogous path integral in this framework is
\begin{equation}\label{eq:mft_trajectory_prob}
\Pr[\rho_t]=\int\mathcal{D}[\rho_s,J_s]\delta\big(\partial_s\rho_s+\partial_xJ_s\big)\mathcal{P}_{\rho_s}[J_s]\mathcal{P}[\rho_0],
\end{equation}
where $\mathcal{P}_{\rho_s}[J_s]\asymp\mathrm{e}^{-\int_0^t\mathrm{d}s\int \mathrm{d}x\tfrac{(J+D(\rho_s)\partial_x\rho_s)^2}{2\sigma(\rho_s)}}$ characterizes the Gaussian fluctuations of the current in terms of two transport coefficients, which for hard-rods take the form \cite{1936_Tonks_The,2005_Lin_From,2025_Grabsch_Exact,2026_Saha_Bottom}: $D(\rho)=\tfrac{D_0}{(1-a\rho)^2}$ and $\sigma(\rho)=2D_0\rho$. Fluctuations in the initial state are characterized in terms of the free energy~\cite{2007_Derrida_Non}
\begin{equation}
\mathcal{F}_{\rm D}[\rho_0]=\int\mathrm{d}x\left\{\mathfrak{m}_0\ln\left(\tfrac{\mathfrak{m}_0}{\bar{\mathfrak{m}}_0}\right)-\left(\mathfrak{m}_0-\bar{\mathfrak{m}}_0\right)\right\},
\end{equation}
where $\mathfrak{m}_0=\tfrac{\rho_0}{1-a\rho_0}$ denotes the renormalized density. Similar to the ballistic case, the generating function of tracer position $z_t$ in \eqref{eq:zt def} is expressed as a path integral with the MFT-action $\mathcal{S}_{\rm D}[\rho,\hat{\rho}]=\mathcal{F}_{\rm D}[\rho_0]+\mathcal{L}_{\rm D}[\rho,\hat{\rho}]$ with
\begin{equation}
\mathcal{L}_{\rm D}=\int_0^t\!\!\!\!\mathrm{d}s\!\!\!\int \!\!\mathrm{d}x\Bigg\{\!\hat{\rho}_s\partial_s\rho_s-\bigg(\frac{\sigma(\rho_s)\partial_x\hat{\rho}_s}{2}-D(\rho_s)\partial_x\rho_s\bigg)\partial_x\hat{\rho}_s\!\!\Bigg\}.
\end{equation}

However, unlike the ballistic case, the presence of noise strongly couples the evolution of $\rho_t$ and $\hat{\rho}_t$ fields, as is evident from the non-linear action. This makes the saddle-point calculations significantly harder, and an explicit solution to recover (\ref{eq:large_dev_one_time_one_tracer},~\ref{eq:large_dev_one_time_two_tracer},~\ref{eq:large_dev_two_time_one_tracer}) is beyond the scope of the present work. That the universality emerges despite these differences makes it all more striking.

\prlsection{Conclusion} In this \emph{Letter}, we report an emergent universality in the one-time statistics of diffusive and ballistic hard-rod gases, despite their fundamentally different dynamics. The universality persists even in the non-stationary setting of a domain-wall initial state \cite{2007_Imamura_Dynamics,2009_Tracy_Asymptotics,2009_Derrida_Current,2009_Derrida_Current2,2017_Doyon_Dynamics,2017_Imamura_Large}, with densities $\rho_-$ and $\rho_+$ to the left and right of the tracer, respectively. Notably, the corresponding annealed single tracer LDFs \cite{Supp_Mat} satisfy a fluctuation symmetry \cite{1993_Evans_Probability,1995_Gallavotti_Dynamical,1998_Kurchan_Fluctuation,1999_Lebowitz_A,1999_Maes_The}, \begin{equation} \phi_{\mathcal{A}}(\xi)-\phi_{\mathcal{A}}(-\xi)=\big(\bar{r}_+-\bar{r}_-\big)\xi, \end{equation} where $r_\pm=\tfrac{\rho_\pm}{1-a\rho_\pm}$ denote the renormalized densities.

Observing the universality in single-file experiments \cite{2005_Lin_From,2004_Kinoshita_Observation,2004_Paredes_Tonks} would be a direct test of our predictions. To this end, a practical extension would be to incorporate more complex inter-particle interactions and spatially varying potentials, settings that remain largely under-explored \cite{2018_Cao_Incomplete,2023_Bagchi_Unusual,2024_Singh_Thermalization,2025_Grabsch_Exact,2026_Saha_Bottom}.
Finally, we leave as exciting questions for the future a formulation of BMFT for the quenched ensemble and a solution of the MFT for the diffusive hard-rod gas.

\prlsection{Acknowledgments} We thank Aitijhya Saha and Abhishek Singh for useful discussions during the initial phase of the project. This research is supported in part by the International Centre for Theoretical Sciences (ICTS) through participation in the program \emph{Hydrodynamics, Fluctuations, and Noise in Quantum and Classical Systems 2025} (code: ICTS/hydrodynamics2025/12). TS and SS are supported by the Department of Atomic Energy, Government of India, under Project Identification Number RTI-4012. JDN and JK are funded by the ERC Starting Grant 101042293 (HEPIQ) and the ANR-22-CPJ1-0021-01. SS gratefully acknowledges the Infosys--TIFR Leading Edge Travel Grant for financial support.

\prlsection{Authors contribution} Soumyabrata Saha and Jitendra Kethepalli contributed equally to this work, performing the calculations and preparing the manuscript. Benjamin Guiselin carried out the numerical simulations and prepared the End Matter. Jacopo De Nardis and Tridib Sadhu designed and supervised the project. All authors contributed to the writing of the manuscript.

\bibliographystyle{apsrev4-2}
\bibliography{references}

\section{End Matter}

\renewcommand{\theequation}{E\arabic{equation}}
\setcounter{equation}{0}
\renewcommand{\thefigure}{E\arabic{figure}}
\setcounter{figure}{0}
\renewcommand{\thetable}{E\arabic{table}}
\setcounter{table}{0}

\subsection{Details on numerical simulations}
We consider the hard‑rod gas of rod length $a=1$, setting the unit for the length scale. We have considered the bare-diffusion coefficient of the hard-rods to be $D_0=0.5$, which sets the unit of time. In all simulations, the gas is prepared at a constant density $\bar{\rho}=0.25$ in the quenched ensemble.

\paragraph{Simulating a single trajectory.}
We simulate a trajectory of $2N+1$ hard-rods with equally spaced initial positions as $ x_0^{(i)}=(i-N-1)/\bar{\rho}\quad \text{for}\quad i=1,\dots, 2N+1, $
so that the tracer ($(N+1)^{\rm th}$ rod) starts at the origin. The hard-rods can be mapped to free particles via the mapping~\cite{1968_Lebowitz_Time}
\begin{equation}
X_0^{(i)}=x_0^{(i)}-\frac{a}{2}\sum_{j=1}^{2N+1}\operatorname{sgn}\left[x_0^{(i)}-x_0^{(j)}\right],
\end{equation}
which gives $X_0^{(i)}=(i-N-1)\left(\tfrac{1}{\bar\rho}-a\right)$ for $i=1,\dots,2N+1$. Each free particle is propagated to time $t$ by the rule
\begin{align}
X_t^{(i)}=X_0^{(i)}+t^{2H}\Delta^{(i)},\label{eqn:update_point_particles_simu}
\end{align}
where $\Delta^{(i)}$ are independent standard Gaussian random variables ($\Delta^{(i)}\sim\mathcal{N}(0,1)$) representing the noise (resp. velocity) for the diffusive (resp. ballistic) gas with $H=\tfrac{1}{4}$ (resp. $H=\tfrac{1}{2}$). This update rule follows directly from the microscopic propagator \eqref{eq:joint_prop_point_part} after mapping it to the free particles.

\paragraph{Position of the tracer.}
The coordinate transformation preserves the order of particles and leaves the median unchanged. Therefore, the position of the tracer at time $t$, $z_t$, is the median of the set $\{X_t^{(i)}\}_{i=1,\dots,2N+1}$. We compute this median efficiently using the QuickSelect algorithm with a median‑of‑three pivot rule.
\paragraph{Time-integrated current across the origin.} We first compute the end position $z_t$ of the tracer as explained above. The sign of the time‑integrated current $n_t$ is then determined by the end position of the tracer: $n_t>0$ if $z_t>0$, and $n_t<0$ otherwise.

In the case $z_t>0$, we must count how many hard-rods initially to the left of the tracer have crossed the origin. To this end, we sort the values $\{X_t^{(i)}\}_{i=1,\dots,N}$ in increasing order using the QuickSort algorithm with a median‑of‑three pivot rule. Traversing this sorted array backwards, we reconstruct the corresponding hard‑rod positions via $x_t^{(i)}=X_t^{(i)}-(N+1-i)a$, for $i=N,\dots,1$, and stop at the first index $i$ where $x_t^{(i)}<0$. The current is then 
\begin{equation}
n_t= N - i + \frac12,
\end{equation}
with the extra $\tfrac{1}{2}$ accounting for the tracer.

In the case $z_t<0$, we instead sort the set $\{X_t^{(i)}\}_{i=N+2,\dots,2N+1}$, reconstruct the hard-rod positions as $x_t^{(i)}=X_t^{(i)}+(i-N-1)a$, for $i=N+2,\dots,2N+1$, stop at the first index $i$ such that $x_t^{(i)}>0$, and obtain
\begin{equation}
n_t= -(i-N-2)-\frac12.
\end{equation}
\paragraph{Two-tracer equal-time correlation function.} At the beginning of the simulation, we select several initial separations $g_0^{(k)}$ between the tracer and a second tagged hard-rod. For each separation, we determine the index $j_k$ of the corresponding rod via
\begin{equation}
g_0^{(k)}=y_0-z_0=x_0^{(j_k)}-x_0^{(N+1)}=x_0^{(j_k)},
\end{equation}
since the tracer is initially at $x_0^{(N+1)}=0$. For each simulated history, we sort the full set $\{X_t^{(i)}\}_{i=1,\dots,2N+1}$ using the QuickSort algorithm with a median-of-three pivot rule. From the sorted array, we extract the end position of the tracer $z_t=X_t^{(N+1)}$ and the position of the second tracer by reverting the coordinate transformation:
\begin{equation}
y_t=X_t^{(j_k)} + (j_k-N-1)a.
\end{equation}

\paragraph{Two-time correlation function of a single tracer.} 
In the ballistic case, we evaluate \eqref{eqn:update_point_particles_simu} at the two times $t_1$ and $t_2$ using the same set of Gaussian random numbers $\{\Delta^{(i)}\}_{i=1,\dots,2N+1}$. This yields the point‑particle positions $X_{t_1}^{(i)}$ and $X_{t_2}^{(i)}$, from which we determine the tracer positions $z_{t_1}$ and $z_{t_2}$ following the procedure described above.

In the diffusive case, we first evaluate \eqref{eqn:update_point_particles_simu} at time $t_1$ and obtain $z_{t_1}$. To evolve the system from $t_1$ to $t_2$, we update the point‑particle coordinates according to
\begin{equation}
X_{t_2}^{(i)}=X_{t_1}^{(i)}+\sqrt{t_2-t_1}{\Delta^{(i)}}',
\end{equation}
where the ${\Delta^{(i)}}'$ are new independent Gaussian random variables with zero mean and unit variance. From this updated configuration, we compute the tracer position at time $t_2$, again using the method described earlier.

\paragraph{Parameter values.} The value of $N$ is chosen large enough to avoid finite-size artefacts. The values of times are chosen large enough to avoid finite-time artefacts.
\begin{itemize}
\item Fig.~\ref{fig:quench_numerics}(a): Ballistic case, tracer position: $N=2^{13}$, $t=100$. Ballistic case, time-integrated current: $N=2^{13}$, $t=100$. Diffusive case, tracer position: $N=2^{13}$, $t=500$. Diffusive case, time-integrated current: $N=2^{13}$, $t=1024$.
\item Fig.~\ref{fig:quench_numerics}(b): Ballistic case: $N=2^{16}$, $t=250$. Diffusive case: $N=2^{13}$, $t=250$.
\item Fig.~\ref{fig:quench_numerics}(c): Ballistic case: $N=2^{16}$ and two values of $t_1=100,\, 250$. Diffusive case: $N=2^{13}$ and two values of $t_1=250,\,500$.
\end{itemize}

\paragraph{Direct simulations.} The data shown in Fig.~\ref{fig:quench_numerics}(b) and Fig.~\ref{fig:quench_numerics}(c) were obtained by straightforward simulations using the numerical procedure described above. For each history, we generate a new set of independent Gaussian random variables $\{\Delta^{(i)}\}$ and compute the desired observable. We repeat this procedure for $s$ independent histories, with $s$ ranging from $10^9$ to $10^{10}$, depending on the observable and the underlying dynamics, to have a sufficient statistical precision.

\paragraph{Importance sampling.} 
To sample the LDF of the tracer position $z_t$ and of the time‑integrated current $n_t$, we use importance‑sampling techniques adapted from \cite{2014_Hartmann_High} (see also \cite{2024_Yerrababu_Dynamical}). We denote by $\mathcal{O}$ the observable whose large‑deviation properties we wish to measure, {\it i.e.,} $\mathcal{O}=\xi=z_t/t^{2H}$ or $\mathcal{O}=j=n_t/t^{2H}$. For a single history, the observable $\mathcal{O}$ is a deterministic function of the set of $2N+1$ Gaussian random variables $\{\Delta^{(i)}\}_{i=1,\dots,2N+1}$.

To efficiently sample the tails of $\mathcal{O}$, we construct a Markov chain in the space of these Gaussian random variables \cite{2025_Hartmann_Numerical}. Starting from a history generated by a given set $\{\Delta^{(i)}\}_{i=1,\dots,2N+1}$, which yields a value $\mathcal{O}$, we propose a trial configuration by replacing $N_{\mathrm{u}}$ randomly chosen entries in the set $\{\Delta^{(i)}\}_{i=1,\dots,2N+1}$ by new Gaussian variables, thus obtaining a trial set $\{{\Delta^{(i)}}'\}_{i=1,\dots,2N+1}$ and the corresponding trial value $\mathcal{O}'$. The trial move is accepted with probability
\begin{equation}
    \min(1,e^{-[W(\mathcal{O}')-W(\mathcal{O})]}),
\end{equation}
where $W$ is a weight function that depends only on $\mathcal{O}$. If the move is accepted, we set $\Delta^{(i)}={\Delta^{(i)}}'$, otherwise the old set is retained. The value of $N_{\mathrm{u}}$ is chosen small enough to maintain a reasonable acceptance rate, while ensuring efficient exploration of configuration space. Its precise value depends on the observable and the underlying dynamics, and is specified below for all cases.

The resulting Markov chain samples the biased distribution
\begin{equation}
{\Pr}^{(W)}(\mathcal{O})=\frac{1}{Z^{(W)}}\Pr(\mathcal{O})e^{-W(\mathcal{O})},
\end{equation}
with $Z^{(W)}$ a normalisation constant. To reconstruct the unbiased probability distribution $\Pr(\mathcal{O})$, we run $m$ independent importance‑sampling simulations in parallel, each using a linear weight $W_a(\mathcal{O})=\beta_a\mathcal{O}$ ($a=1,\dots, m$) with different values of the effective inverse temperature $\beta_a$. Each simulation yields a histogram $\mathcal{N}_a(\mathcal{O})$ of observed $\mathcal{O}$-values, with $s_a=\sum_\mathcal{O} \mathcal{N}_a(\mathcal{O})$ the total number of samples.

The unbiased distribution is obtained through the multi-histogram reweighting method~\cite{1989_Ferrenberg_Optimized,1999_Newman_Monte}:
\begin{equation}
\Pr(\mathcal{O})\mathrm{d}\mathcal{O} = \frac{\sum_{a}\mathcal{N}_a(\mathcal{O})}{\sum_a s_ae^{-W_a(\mathcal{O})}Z_a^{-1}},
\end{equation}
where the normalisation constants $Z_a=Z^{(W_a)}$ are determined self-consistently from
\begin{equation}
Z_a=\sum_\mathcal{O} \frac{\sum_{b}\mathcal{N}_{b}(\mathcal{O})}{\sum_{b} s_{b}e^{W_a(\mathcal{O})-W_{b}(\mathcal{O})}Z_{b}^{-1}}.
\end{equation}
Accurate reconstruction requires that the values of $\beta_a$ are chosen such that histograms from adjacent $\beta_a$’s have substantial overlap.

The LDF $\chi(\mathcal{O})=\phi(\xi),\,\psi(j)$ is then given by
\begin{equation}
\chi(\mathcal{O})=-\frac{1}{t^{2H}}\ln\Pr(\mathcal{O})+\text{constant},
\end{equation}
with the constant chosen so that $\chi$ vanishes at its global minimum. 
The LDF is plotted in Fig. \ref{fig:quench_numerics}(a) and its sampling indeed required rare-events simulations. For instance, in the ballistic case, we are able to sample values of $\mathcal{O}$ such that $\chi(\mathcal{O})\approx6.75$ corresponding to $\tfrac{\Pr(\mathcal{O})}{\Pr(0)}\approx10^{-295}$.

For each $\beta_a$, convergence to the stationary regime is checked by running two independent Markov chains started from different initial sets $\{\Delta^{(i)}\}_{i=1,2\dots,2N+1}$, and verifying that the time series of $\xi$ converge to the same mean value. We then store the final configuration $\{\Delta^{(i)}\}_{i=1,\dots,2N+1}$ and start longer production runs from this configuration to accumulate statistics for the histogram $\mathcal{N}_a(\mathcal{O})$.

\paragraph{Parameter values.} The values of $\beta_a$ and $N_\mathrm{u}$ are as follows.
\begin{itemize}
\item Tracer position, ballistic case: 27 values of $\beta_a$ equally spaced by $25$ between $-325$ and $325$, $N_\mathrm{u}=200$.
\item Tracer position, diffusive case: 25 values of $\beta_a$ equally spaced by $10$ between $-120$ and $120$, $N_\mathrm{u}=600$.
\item Time-integrated current, ballistic case: 17 values of $\beta_a$ equally spaced by $100$ between $-800$ and $800$, $N_\mathrm{u}=200$.
\item Time-integrated current, diffusive case: 13 values of $\beta_a$ equally spaced by $50$ between $-300$ and $300$, $N_\mathrm{u}=1000$.
\end{itemize}

\end{document}

\title{Supplement to `Universal tracer statistics in single-file transport'}

\maketitle

\tableofcontents

\section{Two-tracer statistics: Microscopics}

We follow the procedure outlined in \cite{2015_Krapivksy_Tagged} (see also \cite{1998_Rodenbeck_Calculating,2009_Lizana_Diffusion,2017_Cividini_Tagged}). We start with a gas of $2N+K+1$ rods indexed as $i=\{-N,-N+1,\cdots,N+K\}$ on the real line with position of the $i$-th hard-rod at time $t$ denoted by $x_t^{(i)}$. The two tracers are labeled $i=0$ (initially at $x_0^{(0)}=0$) and $i=K$ (initially at $x_0^{(K)}=g_0$).

The joint-probability of the tracer-positions $x_t^{(0)}\equiv z_t$ and $x_t^{(K)}\equiv y_t$ is given by the marginal subject to the single-file constraint
\begin{align}
P(z_t,y_t|\mathbf{x}_0)=&\left[\int_{-\infty}^{x_t^{(-N+1)}-a}\mathrm{d}x_t^{(-N)}\cdots\int_{-\infty}^{z_t-a}\mathrm{d}x_t^{(-1)}\right]\left[\int_{z_t+a}^{x_t^{(2)}-a}\mathrm{d}x_t^{(1)}\cdots\int_{z_t+(K-1)a}^{y_t-a}\mathrm{d}x_t^{(K-1)}\right]\nonumber\\
&\left[\int_{y_t+a}^{\infty}\mathrm{d}x_t^{(K+1)}\cdots\int_{x_t^{(K+N-1)}+a}^{\infty}\mathrm{d}x_t^{(K+N)}\right]P(\mathbf{x}_t|\mathbf{x}_0)
\end{align}

First, we map this system of hard-rods to an equivalent point-particles representation via the coordinate transformation
\begin{equation}
X_t^{(i)}=x_t^{(i)}-ia
\end{equation}
where $X_t^{(i)}$ denotes the point-particle coordinates. As a result, the tracers' initial (resp. final) positions are $X_0^{(0)}=0$ (resp. $X_t^{(0)}=Z_t$) and $X_0^{(K)}=\Delta_0$ (resp. $X_t^{(K)}=Y_t$) We then arrive at
\begin{subequations}
\begin{align}
P(Z_t,Y_t|\mathbf{X}_0)=&\left[\int_{-\infty}^{X_t^{(-N+1)}}\mathrm{d}X_t^{(-N)}\cdots\int_{-\infty}^{Z_t}\mathrm{d}X_t^{(-1)}\right]\left[\int_{Z_t}^{X_t^{(2)}}\mathrm{d}X_t^{(1)}\cdots\int_{Z_t}^{Y_t}\mathrm{d}X_t^{(K-1)}\right]\nonumber\\
&\left[\int_{Y_t}^{\infty}\mathrm{d}X_t^{(K+1)}\cdots\int_{X_t^{(K+N-1)}}^{\infty}\mathrm{d}X_t^{(K+N)}\right]P(\mathbf{X}_t|\mathbf{X}_0)
\end{align}
where, for the diffusive (stochastic) particles
\begin{equation}
P(\mathbf{X}_t|\mathbf{X}_0)=\sum_{\sigma_i}\prod_{i=-N}^{K+N}\frac{1}{\sqrt{4\pi D_0t}}\exp\bigg[-\frac{\big(X_t^{(i)}-X_0^{(\sigma_i)}\big)^2}{4D_0t}\bigg]
\end{equation}
and for the ballistic (deterministic) particles
\begin{align}
P(\mathbf{X}_t|\mathbf{X}_0)&=\Bigg<\sum_{\sigma_i}\prod_{i=-N}^{K+N}\delta\big(X_{t}^{(i)}-X_{0}^{(\sigma_i)}-\theta^{(\sigma_i)}t\big)\Bigg>_{\theta}\nonumber\\
&=\sum_{\sigma_i}\prod_{i=-N}^{K+N}\frac{1}{\sqrt{4\pi D_0t^2}}\exp\bigg[-\frac{\big(X_{t}^{(i)}-X_{0}^{(\sigma_i)}\big)^2}{4D_0t^2}\bigg]
\end{align}
\end{subequations}

Taking the summation over permutations outside the integral and noting that $X_t^{(i)}$ are dummy variables in each term, we write
\begin{subequations}
\begin{equation}
P(Z_t,Y_t|\mathbf{X}_0)=\sum_{j_1,j_2}P_{j_1,j_2}(Z_t,Y_t|\mathbf{X}_0)
\end{equation}
with each term in the summation
\begin{align}
P_{j_1,j_2}(Z_t,Y_t|\mathbf{X}_0)&=\frac{\mathrm{e}^{-\frac{\left(Z_t-X_0^{(j_1)}\right)^2}{4D_0t^{4H}}}}{\sqrt{4\pi D_0t^{4H}}}\frac{\mathrm{e}^{-\frac{\left(Y_t-X_0^{(j_2)}\right)^2}{4D_0t^{4H}}}}{\sqrt{4\pi D_0t^{4H}}}\Bigg(\prod_{j}\sum_{(\alpha_j,\beta_j)}\Bigg)\Bigg\{\delta_{\sum_{j}\alpha_j\beta_j,-(K-1)}\delta_{\sum_{j}\alpha_j+\beta_j,0}\nonumber\\
&\quad\prod_{j}\bigg[\frac{1+\delta_{\beta_j,0}}{2}\E\bigg(\beta_j\frac{X_0^{(j)}-Z_t}{2\sqrt{D_0}t^{2H}}\bigg)\frac{1+\delta_{\alpha_j,0}}{2}\E\bigg(\alpha_j\frac{X_0^{(j)}-Y_t}{2\sqrt{D_0}t^{2H}}\bigg)\bigg]\Bigg\}
\end{align}
\end{subequations}
where $j\in\{-N,\cdots,K+N\}\setminus\{j_1,j_2\}$ and $(\alpha_j,\beta_j)$ takes the values $\{(1,0),(-1,1),(0,-1)\}$.

In order to extract the large-deviation asymptotic, we only consider the term $(j_1,j_2)=(0,K)$ in the summation. Using an integral representation of the delta-functions and summing over the $(\alpha_j,\beta_j)$-variables, we arrive at
\begin{align}
P_{0,K}(Z_t,Y_t|\mathbf{X}_0)&=\frac{\mathrm{e}^{-\frac{Z_t^2}{4D_0t^{4H}}}}{\sqrt{4\pi D_0t^{4H}}}\frac{\mathrm{e}^{-\frac{\left(Y_t-\Delta_0\right)^2}{4D_0t^{4H}}}}{\sqrt{4\pi D_0t^{4H}}}\int\mathrm{d}B_1\,\mathrm{d}B_2\,\mathrm{e}^{-\frac{B_1-B_2}{2}(K-1)}\prod_{i\in\Omega}\Bigg[\frac{\mathrm{e}^{-\frac{B_1+B_2}{2}}}{2}\E\bigg(\frac{X_0^{(j)}-Z_t}{2\sqrt{D_0}t^{2H}}\bigg)\nonumber\\
&\quad+\frac{\mathrm{e}^{\frac{B_1-B_2}{2}}}{4}\E\bigg(\frac{Z_t-X_0^{(j)}}{2\sqrt{D_0}t^{2H}}\bigg)\E\bigg(\frac{X_0^{(j)}-Y_t}{2\sqrt{D_0}t^{2H}}\bigg)+\frac{\mathrm{e}^{\frac{B_1+B_2}{2}}}{2}\E\bigg(\frac{Y_t-X_0^{(j)}}{2\sqrt{D_0}t^{2H}}\bigg)\Bigg]
\end{align}
where $\Omega=\{-N,\cdots,-1,1,\cdots,K-1,K+1,\cdots,K+N\}$. We further simplify the above expression to obtain
\begin{align}
P_{0,K}(Z_t,Y_t|\mathbf{X}_0)&=\frac{\mathrm{e}^{-\frac{Z_t^2}{4D_0t^{4H}}}}{\sqrt{4\pi D_0t^{4H}}}\frac{\mathrm{e}^{-\frac{\left(Y_t-\Delta_0\right)^2}{4D_0t^{4H}}}}{\sqrt{4\pi D_0t^{4H}}}\int\mathrm{d}B_1\,\mathrm{d}B_2\nonumber\\
&\quad\Bigg\{\prod_{i=-N}^{-1}\bigg[1+\frac{\mathrm{e}^{B_1}-1}{2}\E\bigg(\frac{Z_t-X_0^{(j)}}{2\sqrt{D_0}t^{2H}}\bigg)+\frac{\mathrm{e}^{B_1}(\mathrm{e}^{B_2}-1)}{2}\E\bigg(\frac{Y_t-X_0^{(j)}}{2\sqrt{D_0}t^{2H}}\bigg)\bigg]\nonumber\\
&\quad\prod_{i=1}^{K-1}\bigg[1+\frac{\mathrm{e}^{-B_1}-1}{2}\E\bigg(\frac{X_0^{(j)}-Z_t}{2\sqrt{D_0}t^{2H}}\bigg)+\frac{\mathrm{e}^{B_2}-1}{2}\E\bigg(\frac{Y_t-X_0^{(j)}}{2\sqrt{D_0}t^{2H}}\bigg)\bigg]\nonumber\\
&\quad\prod_{i=K+1}^{K+N}\bigg[1+\frac{\mathrm{e}^{-B_2}(\mathrm{e}^{-B_1}-1)}{2}\E\bigg(\frac{X_0^{(j)}-Z_t}{2\sqrt{D_0}t^{2H}}\bigg)+\frac{\mathrm{e}^{-B_2}-1}{2}\E\bigg(\frac{X_0^{(j)}-Y_t}{2\sqrt{D_0}t^{2H}}\bigg)\bigg]\Bigg\}
\end{align}

\subsection{Annealed initial positions}
The initial positions are drawn randomly from a uniform distribution $\bar{r}$ within an interval $[-L,\Delta_0+L]$ where $\Delta_0$ is the initial separation between the two tracers. Thus, we have
\begin{align}
&P_{0,K}^{\rm ann}(Z_t,Y_t)=\frac{\mathrm{e}^{-\frac{Z_t^2}{4D_0t^{4H}}}}{\sqrt{4\pi D_0t^{4H}}}\frac{\mathrm{e}^{-\frac{\left(Y_t-\Delta_0\right)^2}{4D_0t^{4H}}}}{\sqrt{4\pi D_0t^{4H}}}\int\mathrm{d}B_1\,\mathrm{d}B_2\nonumber\\
&\qquad\qquad\Bigg[1+\frac{\mathrm{e}^{B_1}-1}{2L}\int_{-L}^{0}\mathrm{d}X_0\E\bigg(\frac{Z_t-X_0}{2\sqrt{D_0}t^{2H}}\bigg)+\frac{\mathrm{e}^{B_1}(\mathrm{e}^{B_2}-1)}{2L}\int_{-L}^{0}\mathrm{d}X_0\E\bigg(\frac{Y_t-X_0}{2\sqrt{D_0}t^{2H}}\bigg)\Bigg]^{N}\nonumber\\
&\qquad\qquad\Bigg[1+\frac{\mathrm{e}^{-B_1}-1}{2\Delta_0}\int_{0}^{\Delta_0}\mathrm{d}X_0\E\bigg(\frac{X_0-Z_t}{2\sqrt{D_0}t^{2H}}\bigg)+\frac{\mathrm{e}^{B_2}-1}{2\Delta_0}\int_{0}^{\Delta_0}\mathrm{d}X_0\E\bigg(\frac{Y_t-X_0}{2\sqrt{D_0}t^{2H}}\bigg)\Bigg]^{K-1}\nonumber\\
&\qquad\qquad\Bigg[1+\frac{\mathrm{e}^{-B_2}\left(\mathrm{e}^{-B_1}-1\right)}{2L}\int_{\Delta_0}^{\Delta_0+L}\mathrm{d}X_0\E\bigg(\frac{X_0-Z_t}{2\sqrt{D_0}t^{2H}}\bigg)+\frac{\mathrm{e}^{-B_2}-1}{2L}\int_{\Delta_0}^{\Delta_0+L}\mathrm{d}X_0\E\bigg(\frac{X_0-Y_t}{2\sqrt{D_0}t^{2H}}\bigg)\Bigg]^{N}
\end{align}
For extracting the large-$t$ behavior, we take $N\gg K\gg1$ and $L\gg\Delta_0\gg0$, keeping $N/L=K/\Delta_0=\bar{r}$ finite. We further rescale $Z_t=\tilde{\xi}t^{2H}$, $Y_t=\tilde{\zeta}t^{2H}$, $X_0=xt^{2H}$ and $\Delta_0=\mathcal{L}t^{2H}$ which yields
\begin{subequations}
\begin{equation}
P_{0,K}^{\rm ann}(Z_t,Y_t)\simeq\frac{\mathrm{e}^{-\frac{\xi^2+\left(\zeta-\mathcal{L}\right)^2}{4D_0}}}{4\pi D_0t^{4H}}\int\mathrm{d}B_1\,\mathrm{d}B_2\exp\big[-t^{2H}\mathcal{S}_{\mathcal{A}}(\tilde{\xi},\tilde{\zeta})\big]
\end{equation}
with
\begin{align}
&\mathcal{S}_{\mathcal{A}}(\tilde{\xi},\tilde{\zeta})\simeq-\bar{r}\Bigg\{\int_{-\infty}^{0}\mathrm{d}x\bigg[\frac{\mathrm{e}^{B_1}-1}{2}\E\bigg(\frac{\tilde{\xi}-x}{2\sqrt{D_0}}\bigg)+\frac{\mathrm{e}^{B_1}(\mathrm{e}^{B_2}-1)}{2}\E\bigg(\frac{\tilde{\zeta}-x}{2\sqrt{D_0}}\bigg)\bigg]\nonumber\\
&\qquad\qquad\qquad+\int_{0}^{\mathcal{L}}\mathrm{d}x\bigg[\frac{\mathrm{e}^{-B_1}-1}{2}\E\bigg(\frac{x-\tilde{\xi}}{2\sqrt{D_0}}\bigg)+\frac{\mathrm{e}^{B_2}-1}{2}\E\bigg(\frac{\tilde{\zeta}-x}{2\sqrt{D_0}}\bigg)\bigg]\nonumber\\
&\qquad\qquad\qquad+\int_{\mathcal{L}}^{\infty}\mathrm{d}x\bigg[\frac{\mathrm{e}^{-B_2}(\mathrm{e}^{-B_1}-1)}{2}\E\bigg(\frac{x-\tilde{\xi}}{2\sqrt{D_0}}\bigg)+\frac{\mathrm{e}^{-B_2}-1}{2}\E\bigg(\frac{x-\tilde{\zeta}}{2\sqrt{D_0}}\bigg)\bigg]\Bigg\}
\end{align}
\end{subequations}

In the large-$t$ limit, the integral over $B_{1,2}$ is dominated by the saddle-point which yields the annealed-LDF
\begin{subequations}\label{eq:2_tracer_point_ann_ldf}
\begin{align}
&-\frac{\Phi_\mathcal{A}(\tilde{\xi},\tilde{\zeta})}{\bar{r}}\simeq\int_{-\infty}^{0}\mathrm{d}x\bigg[\frac{\mathrm{e}^{B_1}-1}{2}\E\bigg(\frac{\tilde{\xi}-x}{2\sqrt{D_0}}\bigg)+\frac{\mathrm{e}^{B_1}(\mathrm{e}^{B_2}-1)}{2}\E\bigg(\frac{\tilde{\zeta}-x}{2\sqrt{D_0}}\bigg)\bigg]\nonumber\\
&\qquad\qquad\qquad+\int_{0}^{\mathcal{L}}\mathrm{d}x\bigg[\frac{\mathrm{e}^{-B_1}-1}{2}\E\bigg(\frac{x-\tilde{\xi}}{2\sqrt{D_0}}\bigg)+\frac{\mathrm{e}^{B_2}-1}{2}\E\bigg(\frac{\tilde{\zeta}-x}{2\sqrt{D_0}}\bigg)\bigg]\nonumber\\
&\qquad\qquad\qquad+\int_{\mathcal{L}}^{\infty}\mathrm{d}x\bigg[\frac{\mathrm{e}^{-B_2}(\mathrm{e}^{-B_1}-1)}{2}\E\bigg(\frac{x-\tilde{\xi}}{2\sqrt{D_0}}\bigg)+\frac{\mathrm{e}^{-B_2}-1}{2}\E\bigg(\frac{x-\tilde{\zeta}}{2\sqrt{D_0}}\bigg)\bigg]
\end{align}
where the parameters $B_{1,2}$ are determined from the optimality conditions 
\begin{equation}
\frac{\mathrm{d}\Phi_\mathcal{A}(\tilde{\xi},\tilde{\zeta})}{\mathrm{d}B_1}=0\quad\text{and}\quad\frac{\mathrm{d}\Phi_\mathcal{A}(\tilde{\xi},\tilde{\zeta})}{\mathrm{d}B_2}=0
\end{equation}
\end{subequations}

\subsection{Quenched initial positions}
The initial positions are set deterministically with an equal separation of $\bar{r}^{-1}$ where $K/\Delta_0=\bar{r}$. Thus, we have
\begin{align}
&P_{0,K}^{\rm que}(Z_t,Y_t)=\frac{\mathrm{e}^{-\frac{Z_t^2}{4D_0t^{4H}}}}{\sqrt{4\pi D_0t^{4H}}}\frac{\mathrm{e}^{-\frac{\left(Y_t-\Delta_0\right)^2}{4D_0t^{4H}}}}{\sqrt{4\pi D_0t^{4H}}}\int\mathrm{d}B_1\,\mathrm{d}B_2\nonumber\\
&\qquad\qquad\qquad\exp\Bigg\{\sum_{i=-N}^{-1}\ln\bigg[1+\frac{\mathrm{e}^{B_1}-1}{2}\E\bigg(\frac{Z_t-i/\bar{r}}{2\sqrt{D_0}t^{2H}}\bigg)+\frac{\mathrm{e}^{B_1}(\mathrm{e}^{B_2}-1)}{2}\E\bigg(\frac{Y_t-i/\bar{r}}{2\sqrt{D_0}t^{2H}}\bigg)\bigg]\nonumber\\
&\qquad\qquad\qquad+\sum_{i=1}^{K-1}\ln\bigg[1+\frac{\mathrm{e}^{-B_1}-1}{2}\E\bigg(\frac{i/\bar{r}-Z_t}{2\sqrt{D_0}t^{2H}}\bigg)+\frac{\mathrm{e}^{B_2}-1}{2}\E\bigg(\frac{Y_t-i/\bar{r}}{2\sqrt{D_0}t^{2H}}\bigg)\bigg]\nonumber\\
&\qquad\qquad\qquad+\sum_{i=K+1}^{K+N}\ln\bigg[1+\frac{\mathrm{e}^{-B_2}(\mathrm{e}^{-B_1}-1)}{2}\E\bigg(\frac{i/\bar{r}-Z_t}{2\sqrt{D_0}t^{2H}}\bigg)+\frac{\mathrm{e}^{-B_2}-1}{2}\E\bigg(\frac{i/\bar{r}-Y_t}{2\sqrt{D_0}t^{2H}}\bigg)\bigg]\Bigg\}
\end{align}

For extracting the large-$t$ behavior, we take $N\gg K\gg1$ and $L\gg\Delta_0\gg0$, keeping $N/L=K/\Delta_0=\bar{r}$ finite. We further rescale $Z_t=\tilde{\xi}t^{2H}$, $Y_t=\tilde{\zeta}t^{2H}$, $X_0=xt^{2H}$ and $\Delta_0=\mathcal{L}t^{2H}$, which yields
\begin{subequations}
\begin{equation}
P_{0,K}^{\rm que}(Z_t,Y_t)\simeq\frac{\mathrm{e}^{-\frac{\xi^2+\left(\zeta-\mathcal{L}\right)^2}{4D_0}}}{4\pi D_0t^{4H}}\int\mathrm{d}A\,\mathrm{d}B\exp\big[-t\mathcal{S}_{\mathcal{Q}}(\tilde{\xi},\tilde{\zeta})\big]
\end{equation}
with
\begin{align}
&\mathcal{S}_{\mathcal{Q}}(\tilde{\xi},\tilde{\zeta})\simeq-\bar{r}\Bigg\{\int_{\infty}^{0}\mathrm{d}z\ln\bigg[1+\frac{\mathrm{e}^{B_1}-1}{2}\E\bigg(\frac{\tilde{\xi}-z}{2\sqrt{D_0}}\bigg)+\frac{\mathrm{e}^{B_1}(\mathrm{e}^{B_2}-1)}{2}\E\bigg(\frac{\tilde{\zeta}-z}{2\sqrt{D_0}}\bigg)\bigg]\nonumber\\
&\qquad\qquad\qquad+\int_{0}^{\mathcal{L}}\mathrm{d}z\ln\bigg[1+\frac{\mathrm{e}^{-B_1}-1}{2}\E\bigg(\frac{z-\tilde{\xi}}{2\sqrt{D_0}}\bigg)+\frac{\mathrm{e}^{B_2}-1}{2}\E\bigg(\frac{\tilde{\zeta}-i/\bar{r}}{2\sqrt{D_0}}\bigg)\bigg]\nonumber\\
&\qquad\qquad\qquad+\int_{\mathcal{L}}^{\infty}\mathrm{d}z\ln\bigg[1+\frac{\mathrm{e}^{-B_2}(\mathrm{e}^{-B_1}-1)}{2}\E\bigg(\frac{z-\tilde{\xi}}{2\sqrt{D_0}}\bigg)+\frac{\mathrm{e}^{-B_2}-1}{2}\E\bigg(\frac{z-\tilde{\zeta}}{2\sqrt{D_0}}\bigg)\bigg]\Bigg\}
\end{align}
\end{subequations}

In the large-$t$ limit, the integral over $B_{1,2}$ is dominated by the saddle-point which yields the quenched-LDF
\begin{subequations}\label{eq:2_tracer_point_que_ldf}
\begin{align}
&-\frac{\Phi_{\mathcal{Q}}(\tilde{\xi},\tilde{\zeta})}{\bar{r}}\simeq\int_{\infty}^{0}\mathrm{d}z\ln\bigg[1+\frac{\mathrm{e}^{B_1}-1}{2}\E\bigg(\frac{\tilde{\xi}-z}{2\sqrt{D_0}}\bigg)+\frac{\mathrm{e}^{B_1}(\mathrm{e}^{B_2}-1)}{2}\E\bigg(\frac{\tilde{\zeta}-z}{2\sqrt{D_0}}\bigg)\bigg]\nonumber\\
&\qquad\qquad\qquad+\int_{0}^{\mathcal{L}}\mathrm{d}z\ln\bigg[1+\frac{\mathrm{e}^{-B_1}-1}{2}\E\bigg(\frac{z-\tilde{\xi}}{2\sqrt{D_0}}\bigg)+\frac{\mathrm{e}^{B_2}-1}{2}\E\bigg(\frac{\tilde{\zeta}-i/\bar{r}}{2\sqrt{D_0}}\bigg)\bigg]\nonumber\\
&\qquad\qquad\qquad+\int_{\mathcal{L}}^{\infty}\mathrm{d}z\ln\bigg[1+\frac{\mathrm{e}^{-B_2}(\mathrm{e}^{-B_1}-1)}{2}\E\bigg(\frac{z-\tilde{\xi}}{2\sqrt{D_0}}\bigg)+\frac{\mathrm{e}^{-B_2}-1}{2}\E\bigg(\frac{z-\tilde{\zeta}}{2\sqrt{D_0}}\bigg)\bigg]
\end{align}
where the parameters $B_{1,2}$ are determined from the optimality conditions
\begin{equation}
\frac{\mathrm{d}\Phi_{\mathcal{Q}}(\tilde{\xi},\tilde{\zeta})}{\mathrm{d}B_1}=\frac{\mathrm{d}\Phi_{\mathcal{Q}}(\tilde{\xi},\tilde{\zeta})}{\mathrm{d}B_2}=0
\end{equation}
\end{subequations}

In order to arrive at the corresponding LDFs in the original hard-rod representation, we make the following transformations to \eqref{eq:2_tracer_point_ann_ldf} and \eqref{eq:2_tracer_point_que_ldf}
\begin{equation}
\bar{r}\to\frac{\bar{\rho}}{1-a\bar{\rho}}\quad,\quad\mathcal{L}\to\ell(1-a\bar{\rho})\quad,\quad\tilde{\xi}\to\xi\quad\text{and}\quad\tilde{\zeta}\to\zeta-a\bar{\rho}\ell
\end{equation}

\section{Two-time statistics: Microscopics}
We follow the procedure outlined in \cite{2015_Sadhu_Large} (see also \cite{2015_Sabhapandit_Exact,2025_Chahal_Stochastic}). Let $x_t^{(i)}$ represent the position of the $i$-th hard-rod at time $t$. The initial position of the tracer (labeled $i=0$) is fixed as $x_t^{(0)}=0$. At the times $t_1$ and $t_2$, this tracer is found to be at the positions $x_{t_1}^{(0)}\equiv z_{t_1}$ and $x_{t_2}^{(0)}\equiv z_{t_2}$ respectively. We map this to a system of point particles by the coordinate transformation $x_t^{(i)}=X_t^{(i)}+ia$ such that the position of the $i$-th point particle at time $t$ is $X_t^{(i)}$. Correspondingly, the tracer's position at time $t_1$ (resp. $t_2$) is $X_{t_1}^{(0)}\equiv Z_{t_1}$ (resp. $X_{t_2}^{(0)}\equiv Z_{t_2}$). We will compute the joint distribution of two-time positions within this point-particle representation.

Let $\Gamma^+(Z_{t_j})$ (resp. $\Gamma^-(Z_{t_j})$) denote the number of particles that start from the left (resp. right) of the origin and are found at positions $>Z_{t_j}$ (resp. $<Z_{t_j}$) at time $t_j$. Then, we have the equivalence
\begin{equation}\label{eq:ni_equiv_tracer_posn_part_num}
P\big(X_{t_1}^{(0)}>Z_{t_1},X_{t_2}^{(0)}>Z_{t_2}\big)=P\big(\Gamma^+(Z_{t_1})-\Gamma^-(Z_{t_1})>0,\Gamma^+(Z_{t_2})-\Gamma^-(Z_{t_2})>0\big)
\end{equation}

Taking logarithm on both sides, and then dividing by $t^{2H}$, we write an equivalence of the large-deviation description as
\begin{equation}\label{eq:ni_ldf_reln}
\varphi(\tilde{\xi}_1,\tilde{\xi}_2)=\nu^{0,0}(\tilde{\xi}_1,\tilde{\xi}_2)
\end{equation}
where we have re-scaled the final positions $Z_{t_1}=\tilde{\xi}_1t^{2H}$ and $Z_{t_2}=\tilde{\xi}_2t^{2H}$ and the measurement times $t_1=\tau_1t$ and $t_2=\tau_2t$ with a large-$t$ parameter.

A more convenient way is to compute the cumulant-generating function of $\Gamma^+(Z_{t_j})-\Gamma^-(Z_{t_j})\equiv\mathfrak{n}_j$ instead of the corresponding LDF, which are related via a Legendre transformation
\begin{equation}
\nu^{\mathfrak{n}_1,\mathfrak{n}_2}(\tilde{\xi}_1,\tilde{\xi}_2)=B_1\mathfrak{n}_1+B_2\mathfrak{n}_2-\mu(B_1,B_2,\tilde{\xi}_1,\tilde{\xi}_2)\quad\text{with}\quad\frac{\partial\mu(B_1,B_2,\tilde{\xi}_1,\tilde{\xi}_2)}{\partial B_{1,2}}=\mathfrak{n}_{1,2}
\end{equation}
Setting $\mathfrak{n}_{1,2}=0$ and using \eqref{eq:ni_ldf_reln}, we write the joint 2-time LDF as
\begin{subequations}\label{eq:2_time_tracer_ldf_equiv}
\begin{equation}
\varphi(\tilde{\xi}_1,\tilde{\xi}_2)=-\mu(B_1,B_2,\tilde{\xi}_1,\tilde{\xi}_2)
\end{equation}
with the $B_{1,2}$-parameters determined from
\begin{equation}
\frac{\partial\mu(B_1,B_2,\tilde{\xi}_1,\tilde{\xi}_2)}{\partial B_1}=0\quad\text{and}\quad\frac{\partial\mu(B_1,B_2,\tilde{\xi}_1,\tilde{\xi}_2)}{\partial B_2}=0
\end{equation}
\end{subequations}

Since, the system is now that of non-interacting particles we write the generating-function of $\mathfrak{n}_{1,2}$ by dynamical averaging (stochasticity for diffusive and velocities for ballistic) as
\begin{align}
\big<\mathrm{e}^{B_1\mathfrak{n}_1+B_1\mathfrak{n}_1}\big>_{\text{d}}=\prod_{i}\bigg[&\Big<\mathrm{e}^{B_1\Theta(\mathcal{X}_{t_1}-Z_{t_1})+B_2\Theta(\mathcal{X}_{t_2}-Z_{t_2})}\Big>_{\mathcal{X}_{0}=X_0^{(i)}}\Theta(-X_0^{(i)})\nonumber\\
&+\Big<\mathrm{e}^{-B_1\Theta(Z_{t_1}-\mathcal{X}_{t_1})-B_2\Theta(Z_{t_2}-\mathcal{X}_{t_2})}\Big>_{\mathcal{X}_{0}=X_0^{(i)}}\Theta(X_0^{(i)})\bigg]
\end{align}
where the product is over all the particles in the system.

The average $\left<\ldots\right>_{\mathcal{X}_{0}=X_0^{(i)}}$ is performed using the single-particle propagator for either system. For the stochastic system, the $2$-time propagator reads
\begin{equation}
G\big(X_{t_2}^{(i)},X_{t_1}^{(i)}\big|X_0^{(i)}\big)=\frac{1}{\sqrt{4\pi D_0(t_2-t_1)}}\exp\bigg[-\frac{\big(X_{t_2}^{(i)}-X_{t_1}^{(i)}\big)^2}{4D_0(t_2-t_1)}\bigg]\frac{1}{\sqrt{4\pi D_0t_1}}\exp\bigg[-\frac{\big(X_{t_1}^{(i)}-X_{0}^{(i)}\big)^2}{4D_0t_1}\bigg]
\end{equation}
while for the deterministic system it is given as
\begin{align}
G\big(X_{t_2}^{(i)},X_{t_1}^{(i)}\big|X_0^{(i)}\big)&=\Big<\delta\big(X_{t_2}^{(i)}-X_{t_1}^{(i)}-\theta^{(i)}(t_2-t_1)\big)\delta\big(X_{t_1}^{(i)}-X_{0}^{(i)}-\theta^{(i)}t_1\big)\Big>_{\theta}\nonumber\\
&=\frac{1}{\sqrt{4\pi D_0(t_2-t_1)^2}}\exp\bigg[-\frac{\big(X_{t_2}^{(i)}-X_{t_1}^{(i)}\big)^2}{4D_0(t_2-t_1)^2}\bigg]\frac{1}{t_1}\delta\bigg(\frac{X_{t_1}^{(i)}-X_0^{(i)}}{t_1}-\frac{X_{t_2}^{(i)}-X_{t_1}^{(i)}}{t_2-t_1}\bigg)
\end{align}

\begin{align}
\big<\mathrm{e}^{B_1\mathfrak{n}_1+B_1\mathfrak{n}_1}\big>_{\text{d}}=\prod_i\Bigg\{&\bigg[1+\frac{\mathrm{e}^{B_1}-1}{2}\E\bigg(\frac{Z_{t_1}-X_0^{(i)}}{2\sqrt{D_0}t_1^{2H}}\bigg)+\frac{\mathrm{e}^{B_2}-1}{2}\E\bigg(\frac{Z_{t_2}-X_0^{(i)}}{2\sqrt{D_0}t_2^{2H}}\bigg)\nonumber\\
&\;\;\;+(\mathrm{e}^{B_1}-1)(\mathrm{e}^{B_2}-1)\int_{Z_{t_1}}^{\infty}\mathrm{d}X_{t_1}\int_{Z_{t_2}}^{\infty}\mathrm{d}X_{t_2}\,G\big(X_{t_2},X_{t_1}\big|X_0^{(i)}\big)\bigg]\Theta(-X_0^{(i)})\nonumber\\
&+\bigg[1+\frac{\mathrm{e}^{-B_1}-1}{2}\E\bigg(\frac{X_0^{(i)}-Z_{t_1}}{2\sqrt{D_0}t_1^{2H}}\bigg)+\frac{\mathrm{e}^{-B_2}-1}{2}\E\bigg(\frac{X_0^{(i)}-Z_{t_2}}{2\sqrt{D_0}t_2^{2H}}\bigg)\nonumber\\
&\;\;\;+(\mathrm{e}^{-B_1}-1)(\mathrm{e}^{-B_2}-1)\int_{-Z_{t_1}}^{\infty}\mathrm{d}X_{t_1}\int_{-Z_{t_2}}^{\infty}\mathrm{d}X_{t_2}\,G\big(X_{t_2},X_{t_1}\big|-X_0^{(i)}\big)\bigg]\Theta(X_0^{(i)})\Bigg\}
\end{align}

\subsection{Annealed initial positions}
The CGF of $\mathfrak{n}_{1,2}$ is given as
\begin{equation}
\frac{1}{t^{2H}}\ln\big<\mathrm{e}^{B_1\mathfrak{n}_1+B_2\mathfrak{n}_2}\big>_{\text{d}+\text{p}}\simeq\mu_{\mathcal{A}}(B_1,B_2,\tilde{\xi}_1,\tilde{\xi}_2)
\end{equation}

Using \eqref{eq:2_time_tracer_ldf_equiv}, we obtain
\begin{subequations}\label{eq:2_time_point_ann_ldf}
\begin{align}
-\frac{\varphi_{\mathcal{A}}(\tilde{\xi}_1,\tilde{\xi}_2)}{\bar{r}}\simeq&\int_{-\infty}^{0}\mathrm{d}x\bigg[\frac{\mathrm{e}^{B_1}-1}{2}\E\bigg(\frac{\tilde{\xi}_1-x}{2\sqrt{D_0}\tau_1^{2H}}\bigg)+\frac{\mathrm{e}^{B_2}-1}{2}\E\bigg(\frac{\tilde{\xi}_2-x}{2\sqrt{D_0}\tau_2^{2H}}\bigg)\nonumber\\
&\;\;\;+(\mathrm{e}^{B_1}-1)(\mathrm{e}^{B_2}-1)\int_{\tilde{\xi}_2}^{\infty}\mathrm{d}x_2\int_{\tilde{\xi}_1}^{\infty}\mathrm{d}x_1\,G\big(x_2,x_1\big|x\big)\bigg]\nonumber\\
&+\int_{0}^{\infty}\mathrm{d}x\bigg[\frac{\mathrm{e}^{-B_1}-1}{2}\E\bigg(\frac{x-\tilde{\xi}_1}{2\sqrt{D_0}\tau_1^{2H}}\bigg)+\frac{\mathrm{e}^{-B_2}-1}{2}\E\bigg(\frac{x-\tilde{\xi}_2}{2\sqrt{D_0}\tau_2^{2H}}\bigg)\nonumber\\
&\;\;\;+(\mathrm{e}^{-B_1}-1)(\mathrm{e}^{-B_2}-1)\int_{-\tilde{\xi}_2}^{\infty}\mathrm{d}x_2\int_{-\tilde{\xi}_1}^{\infty}\mathrm{d}x_1\,G\big(x_2,x_1\big|-x\big)\bigg]
\end{align}
with
\begin{equation}
\frac{\mathrm{d}\varphi_{\mathcal{A}}(\tilde{\xi}_1,\tilde{\xi}_2)}{\mathrm{d}B_1}=\frac{\mathrm{d}\varphi_{\mathcal{A}}(\tilde{\xi}_1,\tilde{\xi}_2)}{\mathrm{d}B_2}=0
\end{equation}
\end{subequations}
where we have re-scaled $X_0=xt^{2H}$.

\subsection{Quenched initial positions}
The CGF of $\mathfrak{n}_{1,2}$ is given as
\begin{equation}
\frac{1}{t^{2H}}\Big<\ln\big<\mathrm{e}^{B_1\mathfrak{n}_1+B_2\mathfrak{n}_2}\big>_{\text{d}}\Big>_{\text{p}}\simeq\mu_{\mathcal{Q}}(B_1,B_2,\tilde{\xi}_1,\tilde{\xi}_2)
\end{equation}

Using \eqref{eq:2_time_tracer_ldf_equiv}, we write the LDF as
\begin{subequations}\label{eq:2_time_point_que_ldf}
\begin{align}
-\frac{\varphi_{\mathcal{Q}}(\tilde{\xi}_1,\tilde{\xi}_2)}{\bar{r}}\simeq&\int_{-\infty}^{0}\mathrm{d}x\ln\bigg[1+\frac{\mathrm{e}^{B_1}-1}{2}\E\bigg(\frac{\tilde{\xi}_1-x}{2\sqrt{D_0}\tau_1^{2H}}\bigg)+\frac{\mathrm{e}^{B_2}-1}{2}\E\bigg(\frac{\tilde{\xi}_2-x}{2\sqrt{D_0}\tau_2^{2H}}\bigg)\nonumber\\
&\;\;\;+(\mathrm{e}^{B_1}-1)(\mathrm{e}^{B_2}-1)\int_{\tilde{\xi}_2}^{\infty}\mathrm{d}x_2\int_{\tilde{\xi}_1}^{\infty}\mathrm{d}x_1\,G\big(x_2,x_1\big|x\big)\bigg]\nonumber\\
&+\int_{0}^{\infty}\mathrm{d}x\ln\bigg[1+\frac{\mathrm{e}^{-B_1}-1}{2}\E\bigg(\frac{x-\tilde{\xi}_1}{2\sqrt{D_0}\tau_1^{2H}}\bigg)+\frac{\mathrm{e}^{-B_2}-1}{2}\E\bigg(\frac{x-\tilde{\xi}_2}{2\sqrt{D_0}\tau_2^{2H}}\bigg)\nonumber\\
&\;\;\;+(\mathrm{e}^{-B_1}-1)(\mathrm{e}^{-B_2}-1)\int_{-\tilde{\xi}_2}^{\infty}\mathrm{d}x_2\int_{-\tilde{\xi}_1}^{\infty}\mathrm{d}x_1\,G\big(x_2,x_1\big|-x\big)\bigg]
\end{align}
with
\begin{equation}
\frac{\mathrm{d}\varphi_{\mathcal{Q}}(\tilde{\xi}_1,\tilde{\xi}_2)}{\mathrm{d}B_1}=\frac{\mathrm{d}\varphi_{\mathcal{Q}}(\tilde{\xi}_1,\tilde{\xi}_2)}{\mathrm{d}B_2}=0
\end{equation}
\end{subequations}

In order to arrive at the corresponding LDFs in the original hard-rod representation, we make the following transformations to \eqref{eq:2_time_point_ann_ldf} and \eqref{eq:2_time_point_que_ldf}
\begin{equation}
\bar{r}\to\frac{\bar{\rho}}{1-a\bar{\rho}}\quad,\quad\tilde{\xi}_1\to\xi_1\quad\text{and}\quad\tilde{\xi}_2\to\xi_2
\end{equation}

\section{One-time one-tracer statistics for a step-like initial density profile}

We consider a step density profile \cite{2009_Derrida_Current2,2017_Imamura_Large,2017_Doyon_Dynamics} in the initial state such that the particles are filled with density $\rho_-$ (resp. $\rho_+$) to the left (resp. right) of the tracer (i.e., the origin). Consequently, in the annealed ensemble, the LDF reads
\begin{equation}\label{eq:1_time_1_tracer_step_profile_anneal_ldf}
\phi_{\mathcal{Q}}(\xi)=\Bigg(\sqrt{\frac{\rho_-}{2(1-a\rho_-)}\int_{\xi}^{\infty}\mathrm{d}u\E\left(\frac{u}{\sqrt{4D_0}}\right)}-\sqrt{\frac{\rho_+}{2(1-a\rho_+)}\int_{-\xi}^{\infty}\mathrm{d}u\E\left(\frac{u}{\sqrt{4D_0}}\right)}\Bigg)^2
\end{equation}
which satisfies the Gallavotti-Cohen fluctuation symmetry relation reported in the \emph{Letter}. In the quneched ensemble, the LDF is given by
\begin{subequations}\label{eq:1_time_1_tracer_step_profile_quench_ldf}
\begin{align}
\phi_{\mathcal{Q}}(\xi)=-\frac{\rho_-}{1-a\rho_-}\int_{\xi}^{\infty}\mathrm{d}u\ln\bigg[1+\frac{\mathrm{e}^{B}-1}{2}\E\Big(\frac{u}{\sqrt{4 D_0}}\Big)\bigg]-\frac{\rho_+}{1-a\rho_+}\int_{-\xi}^{\infty}\mathrm{d}u\ln\bigg[1+\frac{\mathrm{e}^{-B}-1}{2}\E\Big(\frac{u}{\sqrt{4 D_0}}\Big)\bigg]
\end{align}
with
\begin{equation}
\frac{\mathrm{d}\phi_{\mathcal{Q}}(\xi)}{\mathrm{d}B}=0
\end{equation}
\end{subequations}

\section{Tracer-current correspondence in quenched ensemble}
Following the steps outlined in \cite{2015_Sadhu_Large}, we determine the two-time current statistics from that of the tracer. Since, the tracer moves from the $x=0$ to $x=z_{t_j}$ in time $t_j$, all particles that were initially located between the origin (the initial tracer position) and $z_{t_j}$ (the final tracer position) must have crossed the position $z_{t_j}$ during the time-period $t_j$. For a quenched ensemble with a uniform initial density $\bar{\rho}$, the number of such particles is simply $\bar{\rho}\left|z_{t_j}\right|$. Thus, the joint probability of finding the tracer at positions $z_{t_1}$ and $z_{t_2}$ at times $t_1$ and $t_2$, respectively, is given by
\begin{equation}
\Pr(z_{t_1},z_{t_2})=\Pr\big(n_{t_1}(z_{t_1})=\bar{\rho}z_{t_1},n_{t_2}(z_{t_2})=\bar{\rho}z_{t_2}\big)
\end{equation}
where the RHS is the joint probability that the time-integrated current across positions $z_{t_1}$ and $z_{t_2}$, up to times $t_1$ and $t_2$, respectively, takes the values $\bar{\rho}z_{t_1}$ and $\bar{\rho}z_{t_2}$.

Using spatial translation invariance of the current statistics, which follows from the uniform initial density, we obtain
\begin{equation}
\Pr(z_{t_1},z_{t_2})=\Pr\big(n_{t_1}(0)=\bar{\rho}z_{t_1},n_{t_2}(0)=\bar{\rho}z_{t_2}\big)
\end{equation}
where the currents are now measured at the origin. Rescaling the observables, we obtain a relation for the corresponding LDFs as
\begin{equation}\label{eq:2_time_ldf_tracer_current_reln}
\varphi_\mathcal{Q}^{(\tau_1,\tau_2)}(\xi_1,\xi_2)=\Psi_\mathcal{Q}^{(\tau_1,\tau_2)}(\bar{\rho}\xi_1,\bar{\rho}\xi_2)
\end{equation}
Note that the $1$-time case, namely $\phi_\mathcal{Q}(\xi)=\psi_\mathcal{Q}(\bar{\rho}\xi)$, is reported in the \emph{Letter} and verified using a rare-event sampling. The relation between the LDFs directly yields the reported relation between the two-time covariances of the tracer and the current.

\section{Quenched-to-annealed two-time covariance}
The annealed two-time covariance can be inferred from the quenched one \cite{1997_Krug_Persistence,2015_Krapivsky_Dynamical,2023_Dandekar_Dynamical}. If the tracer's position is measured at time $t_j$ starting from $t=T\gg0$ instead of $t=0$, we denote the net displacement over the interval $(T,T+t_j)$ as
\begin{equation}
\mathfrak{z}_{t_j}=z_{T+t_j}-z_{T}
\end{equation}
and the corresponding two-time covariance as
\begin{equation}
\big<\mathfrak{z}_{t_1}\mathfrak{z}_{t_2}\big>=\big<z_{T+t_1}z_{T+t_2}\big>_\mathcal{Q}-\big<z_{T+t_1}z_{T}\big>_\mathcal{Q}-\big<z_{T}z_{T+t_2}\big>_\mathcal{Q}+\big<z_{T}^2\big>_\mathcal{Q}
\end{equation}
Using the expression of the two-time quenched covariance, we write
\begin{align}
\frac{\big<\mathfrak{z}_{t_1}\mathfrak{z}_{t_2}\big>}{\kappa}&\simeq\sqrt{(T+t_1)^{4H}+(T+t_2)^{4H}}-\sqrt{(T+t_1)^{4H}+T^{4H}}-\sqrt{T^{4H}+(T+t_2)^{4H}}+\sqrt{2}T^{2H}\nonumber\\
&\quad+t_1^{2H}+t_2^{2H}-|t_1-t_2|^{2H}
\end{align}
where $\kappa=\tfrac{\sqrt{D_0}}{\sqrt{\pi}\bar{r}}$. In the limit of $\tfrac{t_j}{T}\to0$, the above reduces to
\begin{equation}
\big<\mathfrak{z}_{t_1}\mathfrak{z}_{t_2}\big>\simeq\frac{\sqrt{D_0}}{\sqrt{\pi}\bar{r}}\big(t_1^{2H}+t_2^{2H}-|t_1-t_2|^{2H}\big)\equiv\big<z_{t_1}z_{t_2}\big>_\mathcal{A}
\end{equation}
Thus, the annealed two-time covariance follows from the quenched one. This method also extends straightforwardly to the two-time covariance of the time-integrated current, yielding the reported identity in the Letter.

\section{Ballistic Macroscopic fluctuation theory for annealed ensemble}
We use the ballistic macroscopic fluctuation theory \cite{2023_Doyon_Ballistic} as described in the main text to study the statistical properties of different observables. Here, we introduce the key quantities which we use throughout the supplementary. The gas of hard rods is fully characterised by the phase space density $\varrho_t(w, \theta)$, which measures the density around $w$ of particles with velocity close to $\theta$. It evolves via the generalised hydrodynamics (GHD) given by
\begin{align}\label{GHD}
    \partial_t \varrho_t(w, \theta) + \partial_w \big(v_t^{\rm eff}(w, \theta) \varrho_t(w, \theta)\big) = 0~~\text{with}~~v_t^{\rm eff}(w, \theta) = \frac{\theta- a \int d\theta' \theta' \varrho_t(w, \theta')}{1-a \int d\theta' \varrho_t(w, \theta')}.
\end{align}
where the effective velocity represents the cumulative shifts due to collisions.  Despite the interactions, the hard rods admit a mapping to free particles via
\begin{align}\label{mapping_supp}
    r_t(W, \theta) = \frac{\varrho_t(w, \theta)}{1-a \int d\theta' \varrho_t(w, \theta')},~~\text{with}~~    w = W+\frac{a}{2}\int_{-\infty}^{\infty} dW'\int d\theta' \; 
    \mathrm{sign}(W-W')\,r_{t}(W', \theta').
\end{align}
Using the GHD Eq.~\eqref{GHD} one finds that the density $r_t$ evolves deterministically starting from the initial density $r_0(W, \theta) = \varrho_0(w, \theta)/(1-a \int d\theta' \varrho_0(w, \theta'))$ via the Euler-scale GHD
\begin{align}\label{free_ev}
    \partial_t r_t(Z, \theta) + \theta \, \partial_{Z}r_t(Z, \theta) = 0.
\end{align}
Since the trajectory is determined by $r_0$, the initial fluctuations are the sole source of randomness at $t>0$. At equilibrium, the initial profile $r_0$ fluctuates from sample to sample around the typical value $\bar{r}_0$. The probability of observing a particular $r_0$ is then controlled by its free-energy cost and takes the
large-deviation form
\begin{align}
    \mathcal{P}_0[r_0] \asymp \exp\Big(-\mathcal{F}[r_0]\Big), ~~\text{with}~~
    \mathcal{F}[r_0] = \int dZ \int d\theta \;\bigg[r_0 \log 
    \Big(\frac{r_0}{\bar{r}_0}\Big)-(r_0-\bar{r}_0)\bigg],
\end{align}
which penalizes deviations from $\bar{r}_0$. Since the dynamics is deterministic, the probability weight of any trajectory is given by
\begin{align}\label{prho_t}
    \mathcal{P}_t[r_t] = \mathcal{P}_0[r_0]\;\delta\big[\partial_s r_s 
    + \theta \, \partial_{Z}r_s\big].
\end{align}
Observables that can be represented as a functional of phase space density can now be studied by computing their cumulant generating function via the BMFT formalism in the free particle coordinates, following the method proposed in Ref.~\cite{2026_Kethepalli_Ballistic}.

\subsection{The $n$ time statistics of the $k^{\rm th}$ rod}

We can characterise the dynamics of the $k^{\rm th}$ rod by studying the joint probability 
distribution $\text{Pr}(\{z^{(k)}_{t_i}-z^{(k)}_0=\xi_i\}_{i = 1}^n)$ of its displacements $z^{(k)}_{t_i}-z^{(k)}_0=\xi_i$ at $n$ successive times, $0<t_1<t_2 <\cdots <t_{n-1}<t_n=t$. In the free particle coordinates, $Z_i^{(k)}\equiv Z^{(k)}_{t_i}$ represents the position of the free particle corresponding to the tracer rod located at $z^{(k)}_i \equiv z^{(k)}_{t_i}$ and they are related by
\begin{align}\label{cont_map}
    z_i[Z_i, r_{t_i}] = Z_{i}+\frac{a}{2}\int_{-\infty}^{\infty} dZ'\int d\theta \; 
    \mathrm{sign}(Z_{i}-Z')\,r_{t_i}(Z', \theta).
\end{align}
Here, the position of the point particle is fixed implicitly via the single-file constraint
\begin{align}\label{single-file}
    \mathscr{N}[Z^{(k)}_i, r_{t_i}] = \int_{-\infty}^{Z^{(k)}_i}dZ \int d\theta \; 
    r_{t_i}(Z, \theta) = k,
\end{align}
which ensures there are exactly $k$ particles to the left at any time $t_i$.

To obtain the joint distribution 
$\mathscr{P}_n(\{z^{(k)}_{t_i}-z^{(k)}_0\}_{i=1}^n)$ we analyse its  
cumulant generating function
\begin{align}
    \log \mathcal{G}_n(\{\lambda_i\}_{i=1}^n) = \log\left\langle \exp\Big(\sum_{i=1}^n 
    \lambda_i \big(z^{(k)}_{i}-z^{(k)}_{0}\big)\Big)\right\rangle,
\end{align}
where the average is taken over initial profiles $r_0$ weighted by $\mathcal{P}_0[r_0]$, 
and the displacements $z^{(k)}_i - z^{(k)}_0$ are expressed through the mapping 
Eq.~\eqref{cont_map}. This average can be expressed as a path integral over the phase-space density field $r_t$, with trajectories weighted by $\mathcal{P}_t[r_t]$, as
\begin{align}\label{G_1}
    \mathcal{G}_n(\{\lambda_i\}_{i=1}^n) = \int \mathcal{D}[r_s] \exp\!\Big(\sum_{i=1}^n 
    \lambda_i \big(z^{(k)}_{i}[Z_i^{(k)}, r_{t_i}]-z^{(k)}_{0}[Z_0^{(k)}, r_{0}]\big)\Big) \mathcal{P}_t[r_t].
\end{align}
Substituting Eq.~\eqref{prho_t} in Eq.~\eqref{G_1} and using the Fourier representation of the delta functional, the path integral becomes
\begin{align}\label{G_2}
    \mathcal{G}_n(\{\lambda_i\}_{i=1}^n) &= \int \mathcal{D}[r_s] \mathcal{D}[\hat{r}_s]\exp\Big(-S[r_s, \hat{r}_s]\Big),~~\text{where}~~\\
    S[r_s, \hat{r}_s] &= \mathcal{F}[r_0]-\sum_i \lambda_i \big(z^{(k)}_{i}[Z_i^{(k)}, r_{t_i}]-z^{(k)}_{0}[Z_0^{(k)}, r_{0}]\big) + \int_0^{t}ds\int dZ' \int d\theta' \hat{r}_{s} \Big(\partial_s r_s 
    + \theta'  \partial_{Z'}r_s\Big). 
\end{align}
To extract the large-$T$ behaviour, we rescale position and time as $Z = t\mathscr{Z}$ 
and $s = t\tau$, under which the action scales as $ S[r_s, \hat{r}_s] = t \mathcal{S}[q_{\tau}, p_{\tau}]$ where
\begin{align}
 \label{scaled_action} \mathcal{S}[q_{\tau}, p_{\tau}]&=\mathfrak{F}[q_0]-\sum_i \lambda_i~\big(\mathfrak{z}_{i}[\mathscr{Z}_i, q_{\tau_i}]-\mathfrak{z}_{0}[\mathscr{Z}_0, q_{0}]\big)+\int_0^{1} d \tau \int d \mathscr{Z}\int d\theta p_{\tau}\Big[\partial_{\tau} q_{\tau} + \theta \partial_{\mathscr{Z}} q_{\tau}\Big].
\end{align}
Here, the scaled density $q_{\tau}(\mathscr{Z}, \theta)$ and its conjugate $p_{\tau}(\mathscr{Z}, \theta)$ are functions of rescaled variables
\begin{align}\label{rescaling}
    &q_{\tau}(\mathscr{Z}, \theta) = r_s(Z, \theta),~~\bar{q}_{\tau}(\mathscr{Z}, \theta) = \bar{r}_s(Z, \theta)~~\text{and}~~  p_{\tau}(\mathscr{Z}, \theta) = \hat{r}_s(Z, \theta)\\
    &\text{with}~~ \tau = \frac{s}{t}~~\text{and}~~\mathscr{Z} = \frac{Z}{t}.
\end{align}
In Eq.~\eqref{scaled_action}, the scaled position of the tagged particle and scaled free energy are
\begin{align}\label{scaled-position}
&\mathfrak{z}_{i}[\mathscr{Z}_i, q_{\tau_i}] =\frac{z_{i}[Z_i, r_{t_i}]}{t}= \mathscr{Z}_{i}+\frac{a}{2}\int_{-\infty}^{\infty} d\mathscr{Z}'\int d\theta \; 
    \mathrm{sign}(\mathscr{Z}_{i}-\mathscr{Z}')\,q_{\tau_i}(\mathscr{Z}', \theta),\\
\label{scaled-free-energy}    &\mathfrak{F}[q_0] = \frac{\mathcal{F}[r_0]}{t} =\int d\mathscr{Z} \int d\theta \;\bigg[q_0 \log 
    \Big(\frac{q_0}{\bar{q}_0}\Big)-(q_0-\bar{q}_0)\bigg].
\end{align}
As the action scales linearly with $t$ and the path integral is dominated by the saddle point of $\mathcal{S}$ in the large-$t$ limit. Varying $\mathcal{S}$ with respect to $q_\tau$, $p_\tau$, and their boundary values yield the following saddle-point equations
\begin{align}\label{peqn_n_t}
    \frac{\delta \mathcal{S}[q_{\tau}, p_{\tau}]}{\delta q_{\tau}(\mathscr{Z}, \theta)} &= 0 = \partial_{\tau} p_{\tau} + \theta \partial_{\mathfrak{Z}} p_{\tau}-\sum_{i \neq n} \lambda_{i} \bigg[\frac{\Theta(\mathscr{Z}_{i}-\mathscr{Z}) }{q_{\tau_i}(\mathscr{Z}_{i})}+\frac{a}{2}\bigg] \delta(\tau-\tau_i)\\
    \label{pic_n_t} \frac{\delta \mathcal{S}[q_{\tau}, p_{\tau}]}{\delta q_0(\mathscr{Z}, \theta)} &= 0= p_0+\sum_{i } \lambda_i\bigg[\frac{\Theta(\mathscr{Z}_{0}-\mathscr{Z}) }{q_{0}(\mathscr{Z}_{0})}+\frac{a}{2}\bigg]-\log\bigg(\frac{q_0(\mathscr{Z}, \theta)}{\bar{q}_0(\mathscr{Z}, \theta)}\bigg)\\
   \label{pfc_n_t} \frac{\delta \mathcal{S}[q_{\tau}, p_{\tau}]}{\delta  q_{1}(\mathscr{Z}, \theta)} &= 0=p_{1}+\lambda_{n} \bigg[\frac{\Theta(\mathscr{Z}_{n}-\mathscr{Z}) }{q_{1}(\mathscr{Z}_{n})}+\frac{a}{2}\bigg]\\
    \frac{\delta \mathcal{S}[q_{\tau}, p_{\tau}]}{\delta p_{\tau}(\mathscr{Z}, \theta)} &= 0 =  \partial_{\tau} q_{\tau} + \theta \partial_{\mathscr{Z}} q_{\tau},\label{qeqn_n_t}
\end{align}
where note that $\tau_n = t_n/t = 1$. Solving the saddle point using the characteristic method~\cite{2026_Kethepalli_Ballistic}, we get
\begin{align}\label{saddle_n_t}
    q_0^*(\mathscr{Z}, \theta) =\bar{q}_0(\mathscr{Z}, \theta) \exp\Bigg(-\sum_{i} \bigg[A_i \Theta(\mathscr{Z}_{0}-\mathscr{Z})+B_i\Theta(\mathscr{Z}_{i}-\tau_i\theta-\mathscr{Z})\bigg]\Bigg),
\end{align}
where the parameters 
\begin{align}
    A_i = -\frac{\lambda_i}{q_{0}(\mathscr{Z}_0)}~~\text{and}~~B_i = \frac{\lambda_i}{q_{\tau_i}(\mathscr{Z}_i)}.
\end{align}
The saddle-point density \eqref{saddle_n_t} contains $2n$ parameters $\{A_i, B_i\}$, but they are not independent. The relations $A_i/\lambda_i = A_j/\lambda_j$ for all $i,j$ imply $A_i = A\,\lambda_i$, reducing the $n$ parameters $\{A_i\}$ to a single parameter $A$, so that the total number of relevant parameters reduces to $n+1$. The condition $ \lim_{Z \to \infty} q_0^*(Z,\theta) = \lim_{Z \to \infty} \bar q_0(Z,\theta) $ then gives the constraint $\sum_i (A_i + B_i)=0$, leaving $n$ relevant parameters. These remaining parameters are determined by the $n$ single-file constraints together with the stationarity conditions obtained by optimizing the action with respect to $A$ and $B_i$, providing a closed set of $n$ equations that fully determine the saddle-point solution. Using the relation between these parameters, the scaled cumulant generating function reduces to a simplified form
\begin{align}
    \mu_n(\{\lambda_i\}_{i=1}^n) = \sum_i \lambda_i~\big(\mathfrak{z}_{i}[\mathscr{Z}_i, q_{\tau_i}^*]-\mathfrak{z}_{0}[\mathscr{Z}_0, q_{0}^*]\big)+\int d\mathscr{Z}\int d\theta \big(q_0^*-\bar{q}_0\big).
\end{align}
Using the Legendre transform, we find the rate function is given by
\begin{align}
    \phi_n\big(\{\mathfrak{z}_i-\mathfrak{z}_0\}_{i=1}^n\big) = -\int d\mathscr{Z}\int d\theta \big(q_0^*-\bar{q}_0\big),
\end{align}
where the saddle point density is given in Eq.~\eqref{saddle_n_t}.

For a general $n$ and $\bar{q}_0$, proceeding further analytically is difficult. However, for the piecewise homogeneous initial profile $\bar{\varrho}_0(w, \theta) = \bar{\varrho}_{+}\Theta(Z)\,h(\theta) + \bar{\varrho}_{-}\Theta(-Z)\,h(\theta)$ can be mapped to the scaled free coordinates as $\bar{q}_0(Z, \theta) = \bar{r}_{+}\Theta(Z)\,h(\theta) + \bar{r}_{-}\Theta(-Z)\,h(\theta)$, where $r_{\pm} = \varrho_{\pm}/(1-a\varrho_{\pm})$ and we can explicitly solve for $n=1$ and $n=2$.  

The rate function for $n=2$ splits into two branches depending on the ordering of the rescaled velocities $(\mathfrak{z}_i - \mathfrak{z}_0)/\tau_i$: 
\begin{align}\label{rate1}
  -\phi_2(\mathfrak{z}_1-\mathfrak{z}_0, \mathfrak{z}_2-\mathfrak{z}_0) =\begin{cases}
      &~~r_{+}(e^{-A_1}-1)~G_c^{10} +  \notag r_{+}(e^{-A_2}-1)~G_c^{20} \\
&+ r_{+}(e^{-A_1}-1)(e^{-A_2}-1)~(G_c^{20}-G_c^{21}) \\ \notag
&+ r_{-}(e^{A_1}-1)~G^{10} + r_{-}(e^{A_2}-1)~G^{20} \\ \notag
&+ r_{-}(e^{A_1}-1)(e^{A_2}-1)~G^{10} ~~\text{for}~~ \frac{\mathfrak{z}_1-\mathfrak{z}_0}{\tau_1}>\frac{\mathfrak{z}_2-\mathfrak{z}_0}{\tau_2}\\
      & r_{+}(e^{-A_1}-1)~G_c^{10} +  \notag r_{+}(e^{-A_2}-1)~G_c^{20} \\
&+ r_{+}(e^{-A_1}-1)(e^{-A_2}-1)~G_c^{10} \\ \notag
&+ r_{-}(e^{A_1}-1)~G^{10} + r_{-}(e^{A_2}-1)~G^{20} \\ \notag
&+ r_{-}(e^{A_1}-1)(e^{A_2}-1)~(G^{20}-G^{21})~~ ~~\text{for}~~ \frac{\mathfrak{z}_1-\mathfrak{z}_0}{\tau_1}<\frac{\mathfrak{z}_2-\mathfrak{z}_0}{\tau_2}
  \end{cases}
\end{align}
where the function $G^{ij}$ and $G_c^{ij}$ are 
\begin{align}
    &G_c^{ij}=G_c^{ji} = (\tau_i-\tau_j)F_c\Big(\frac{\mathscr{Z}_i-\mathscr{Z}_j}{\tau_i-\tau_j}\Big)~~~\text{with}~~~F_c(x) = \int_{-x}^{\infty} d\theta \int_{\theta}^{\infty} ds~h(s)\\
    &G^{ij}=G^{ji} = (\tau_i-\tau_j)F\Big(\frac{\mathscr{Z}_i-\mathscr{Z}_j}{\tau_i-\tau_j}\Big)~~~\text{with}~~~F(x)= \int_{x}^{\infty} d\theta \int_{\theta}^{\infty} ds~h(s).
\end{align}
Here note that $\mathscr{Z}_i-\mathscr{Z}_0 = \mathfrak{z}_i - \mathfrak{z}_0$ as it is the displacement. The parameters in Eq.~\eqref{rate1} satisfy the following self consistency relation for (a) $\frac{\mathfrak{z}_1-\mathfrak{z}_0}{\tau_1}<\frac{\mathfrak{z}_2-\mathfrak{z}_0}{\tau_2}$
\begin{align}
e^{2 A_1}&=\frac{r_{+}}{r_{-}}\frac{G_c^{10}+G_c^{10}\big(e^{-A_2}-1\big)}{G^{10} +  (e^{A_2}-1)(G^{20}-G^{21})},\\
e^{2 A_2} &= \frac{r_{+}}{r_{-}} \frac{G_c^{20}+G_c^{10}\Big(e^{-A_1}-1\Big)}{G^{20}+(e^{A_1}-1)(G^{20}-G^{21})}.
\end{align}
and for (b) $\frac{\mathfrak{z}_1-\mathfrak{z}_0}{\tau_1}>\frac{\mathfrak{z}_2-\mathfrak{z}_0}{\tau_2}$
\begin{align}\label{self-consistency1}
e^{2 A_1} &= \frac{r_{+}}{r_{-}}\frac{G_c^{10}+\big(G_c^{20}-G_c^{21}\big)\Big(e^{-A_2}-1\Big)}{G^{10}+G^{10}( e^{A_2}-1)} ,  \\
   e^{2 A_2} &=\frac{r_{+}}{r_{-}}\frac{G_c^{20} + \big(G_c^{20}-G_c^{21}\big)\Big(e^{-A_1}-1\Big)}{G^{20} + G^{10}(e^{A_1}-1)}.
\end{align}

We can now compute the rate function for $n=1$ by substituting $A_2 = 0$ in Eq.~\eqref{rate1}, which returns 
\begin{align}\label{rate0}
    -\phi_1(\mathfrak{z}_1-\mathfrak{z}_0) &= r_{+}(e^{-A_1}-1)~G_c^{10} 
+ r_{-}(e^{A_1}-1)~G^{10},\\
&=\Big(\sqrt{r_{+}G_c^{10}}-\sqrt{r_{-}G^{10}}\Big)^{2}.
\end{align}
where we used Eq.~\eqref{self-consistency1} with $A_2 = 0$ which gives
\begin{align}
e^{A_1} &= \sqrt{\frac{r_{+}}{r_{-}}\frac{G_c^{10}}{G^{10}} }.
\end{align}
To 
Both Eq.~\eqref{rate1} and ~\eqref{rate0} match the results in the maintext for $\bar{r}_+ = \bar{r}_{-} = \bar{r}$ where $\bar{r} = \bar{\rho}/(1-a\bar{\rho})$.

\subsection{The $n$ rod statistics at time $t$}
We study the joint probability distribution $\mathscr{P}_n(\{z_{t}^{(i)}-z_0^{(i)}\}_{i=1}^n)$ of the displacements of $n$ tagged rods at time $t$. We use the same approach as discussed above to compute the cumulant generating function
\begin{align}\label{gen_n_tag}
    \log G_n(\{\lambda_i\}_{i=1}^n) = \log\left\langle \exp\Big(\sum_{i=1}^n 
    \lambda_i \big(z^{(i)}_{t}-z^{(i)}_{0}\big)\Big)\right\rangle,
\end{align}
where the average is taken over initial profiles $r_0$ weighted by $\mathcal{P}_0[r_0]$, 
and the displacements $z^{(i)}_t - z^{(i)}_0$ are expressed through the mapping 
Eq.~\eqref{cont_map}.

The average in Eq.~\eqref{gen_n_tag} can be expressed as a path integral over the phase-space density $r_t$ and its conjugate $\hat{r}_t$ given by
\begin{align}\label{G_2_n}
    G_n(\{\lambda_i\}_{i=1}^n) &= \int \mathcal{D}[r_t] \mathcal{D}[\hat{r}_t]\exp\Big(-S[r_t, \hat{r}_t]\Big)~\text{where}\\
    S[r_t, \hat{r}_t] &= \mathcal{F}[r_0]-\sum_i \lambda_i \big(z^{(i)}_{t}[Z_t^{(i)}, r_{t}]-z^{(i)}_{0}[Z_0^{(i)}, r_{0}]\big) + \int_0^{T}dt\int dZ' \int d\theta' \hat{r}_{t} \Big(\partial_t r_t 
    + \theta'  \partial_{Z'}r_t\Big). 
\end{align}
To extract the large-$T$ behaviour, we rescale position and time as $Z = T\mathscr{Z}$  and $t = T\tau$, under which the action scales as
\begin{align}
    S[r_t, \hat{r}_t] &= T \mathcal{S}[q_{\tau}, p_{\tau}]\\
    \label{scaled_action_n}\mathcal{S}[q_{\tau}, p_{\tau} ]&=\mathfrak{F}[q_0]-\sum_i \lambda_i~\big(\mathfrak{z}_{1}^{(i)}[\mathscr{Z}_{1}^{(i)}, q_{1}]-\mathfrak{z}_{0}^{(i)}[\mathscr{Z}_{0}^{(i)}, q_{0}]\big) +\int_0^{1} d \tau \int d \mathscr{Z}\int d\theta p_{\tau}\Big[\partial_{\tau} q_{\tau} + \theta \partial_{\mathscr{Z}} q_{\tau}\Big],
\end{align}
where $q_{\tau}(\mathscr{Z}, \theta)$ and $p_{\tau}(\mathscr{Z}, \theta)$ are functions of rescaled variables same as Eq.~\eqref{rescaling} and the relation between the scaled positions of the tagged particles $\mathfrak{z}_{1}^{(i)}$  and free particle $\mathscr{Z}_{1}^{(i)}$ is Eq.~\eqref{scaled-position} while the scaled free energy is given by Eq.~\eqref{scaled-free-energy}.

As the action scales linearly with $t$, the path integral is dominated by the saddle point of $\mathcal{S}$ in the large-$t$ limit. Varying $\mathcal{S}$ with respect to $q_\tau$, $p_\tau$, and their boundary values yield the following saddle-point equations
\begin{align}\label{peqn_n}
    \frac{\delta \mathcal{S}[q_{\tau}, p_{\tau}]}{\delta q_{\tau}(\mathfrak{Z}, \theta)} &= 0 = \partial_{\tau} p_{\tau} + \theta \partial_{\mathscr{Z}} p_{\tau}\\
    \label{pic_n} \frac{\delta \mathcal{S}[q_{\tau}, p_{\tau}]}{\delta q_0(\mathscr{Z}, \theta)} &= 0= p_0+\sum_{i } \lambda_i\bigg[\frac{\Theta(\mathscr{Z}_{0}-\mathscr{Z}) }{q_{0}(\mathscr{Z}_{0})}+\frac{a}{2}\bigg]-\log\bigg(\frac{q_0(\mathscr{Z}, \theta)}{\bar{q}_0(\mathscr{Z}, \theta)}\bigg)\\
   \label{pfc_n} \frac{\delta \mathcal{S}[q_{\tau}, p_{\tau}]}{\delta  q_{1}(\mathscr{Z}, \theta)} &= 0=p_{1}+\sum_{i } \lambda_{i} \bigg[\frac{\Theta(\mathscr{Z}_1^{(i)}-\mathscr{Z}) }{q_{1}(\mathscr{Z}_1^{(i)})}+\frac{a}{2}\bigg]\\
    \frac{\delta \mathcal{S}[q_{\tau}, p_{\tau}]}{\delta p_{\tau}(\mathscr{Z}, \theta)} &= 0 =  \partial_{\tau} q_{\tau} + \theta \partial_{\mathscr{Z}} q_{\tau}.\label{qeqn_n}
\end{align}
Solving the saddle point, we get
\begin{align}\label{saddle_n}
    q_0^*(\mathscr{Z}, \theta) =\bar{q}_0(\mathscr{Z}, \theta) \exp\Bigg(-\sum_{i} \bigg[C_i \Theta(\mathscr{Z}_0^{(i)}-\mathscr{Z})+D_i\Theta(\mathscr{Z}_1^{(i)}-\theta-\mathscr{Z})\bigg]\Bigg),
\end{align}
where the parameters are
\begin{align}
    C_i = -\frac{\lambda_i}{q_{0}^*(\mathscr{Z}_0^{(i)})}~~\text{and}~~D_i = \frac{\lambda_i}{q_{1}^*(\mathscr{Z}_1^{(i)})}.
\end{align}
The saddle-point density \eqref{saddle_n} contains $2n$ parameters $\{C_i, D_i\}$, but they are not independent. The parameters $C_i$ and $D_i$ are determined by extremizing the cumulant generating function with respect to $C_i$ and $D_i$, which yields a coupled set of equations for each $i$. These equations consistently admit a solution, $C_i + D_i = 0$, thereby reducing the number of independent parameters to $n$. This can be self-consistently checked by computing the marginal density of Eq.~\eqref{saddle_n} at $\mathscr{Z}_0^{(i)}$ and $\mathscr{Z}_1^{(i)}$ which gives $q_{0}^*(\mathscr{Z}_0^{(i)}) =q_{1}^*(\mathscr{Z}_1^{(i)})$ equivalently $C_i+D_i = 0$. The remaining parameters are fixed by the $n$ single-file constraints \eqref{single-file}. We can hence express the cumulant generating function as
\begin{align}
    \mu_n(\{\lambda_i\}_{i=1}^n) = \sum_i \lambda_i~\big(\mathfrak{z}_{1}^{(i)}[\mathscr{Z}_{1}^{(i)}, q_{1}]-\mathfrak{z}_{0}^{(i)}[\mathscr{Z}_{0}^{(i)}, q_{0}]\big)+\int d\mathscr{Z}\int d\theta \big(q_0^*-\bar{q}_0\big).
\end{align}
Using the Legendre transform, we find the rate function is given by
\begin{align}
    \Phi_{n}\big(\{\mathfrak{z}_1^{(i)}-\mathfrak{z}_0^{(i)}\}_{i=1}^n\big) = -\int d\mathscr{Z}\int d\theta \big(q_0^*-\bar{q}_0\big),
\end{align}
where the saddle point density is given in Eq.~\eqref{saddle_n}.  

For a general $n$ and $\bar{q}_0$, proceeding further analytically is difficult. However for the piecewise homogeneous initial profile $\bar{q}_0 = \bar{r}_{+}\Theta(Z)\,h(\theta) + \bar{r}_{-}\Theta(-Z)\,h(\theta) $, we can explicitly solve for $n=1$ and $n=2$.  The rate function for $n=1$ is exactly Eq.~\eqref{rate0} since it is the single tracer statistics. The rate function for $n=2$ is 
\begin{align}\label{raten}
   -\Phi_2(\mathfrak{z}_1^{(1)}-\mathfrak{z}_0^{(1)}, \mathfrak{z}_1^{(2)}-\mathfrak{z}_0^{(2)}) 
    &=r_{-}H^{11}\big(1-e^{C_1}\big) + r_{+}H_c^{11}\big(1-e^{-C_1}\big) -r_{-}H^{21} \big(1-e^{C_1}\big)\big(1-e^{C_2}\big) \notag \\
    & +  r_{-}H^{22}  \big(1-e^{C_2}\big)  +r_{+}H^{22}_c   \big(1-e^{-C_2}\big)- r_{+}H_c^{12}  \big(1-e^{-C_1}\big)\big(1-e^{-C_2}\big).
\end{align}
where the functions are
\begin{align}
    &H_c^{ij}= F_c\Big(\mathscr{Z}_1^{(i)}-\mathscr{Z}_0^{(j)}\Big)~~~\text{with}~~~F_c(x) = \int_{-x}^{\infty} d\theta \int_{\theta}^{\infty} ds~h(s)\\
    &H^{ij}= F\Big(\mathscr{Z}_1^{(i)}-\mathscr{Z}_0^{(j)}\Big)~~~\text{with}~~~F(x)= \int_{x}^{\infty} d\theta \int_{\theta}^{\infty} ds~h(s).
\end{align}
Here the interacting and free coordinates are related using the mapping Eq.~\eqref{cont_map} which gives $\mathscr{Z}_{0/1}^{(1)} = \mathfrak{z}_{0/1}^{(1)} $, $\mathscr{Z}_0^{(2)} = \mathfrak{z}_0^{(2)}/(1+a \bar{r}_{+})$ and $ \mathscr{Z}_1^{(2)} = \mathfrak{z}_1^{(2)} - a \bar{r}_{+} \mathscr{Z}_0^{(2)}$. The parameters $C_1, C_2$ are obtained from the single-file constraint for each particle Eq.~\eqref{single-file}
\begin{align}
&e^{2C_2}  =  \frac{r_{+}}{r_{-}}\frac{H_c^{22} +  \big(e^{-C_1}-1\big)H_c^{12}}{H^{22}+(e^{C_1}-1)  H^{21}  },\label{i-equation1}\\
&e^{2C_1} =\frac{r_{+}}{r_{-}} \frac{ H_c^{11} +  \big(e^{-C_2}-1\big)H_c^{12}}{H^{11}+ (e^{C_2}-1) H^{21} }  .\label{j-equation1}
\end{align}
The result in Eq.~\eqref{raten} matches the main text for $\bar{r}_{+} = \bar{r}_{-} = \bar{r}$.

\bibliographystyle{apsrev4-2}
\bibliography{references}